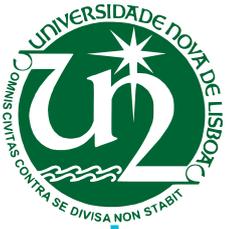

**Ana Catarina Malhado Ribeiro**

MSc Student

# Invariant-Driven Automated Testing

Dissertation submitted in partial fulfillment
of the requirements for the degree of

Master of Science in
**Computer Science and Informatics Engineering**

Adviser: Carla Ferreira, Associate Professor,
NOVA University of Lisbon

Examination Committee

Chairperson: António Ravara, Associate Professor, NOVA University of Lisbon
Raporteur: Jácome Cunha, Assistant Professor, University of Minho
Member: Carla Ferreira, Associate Professor, NOVA University of Lisbon

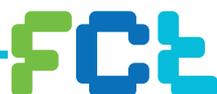

**February, 2021**

**Invariant-Driven Automated Testing**

# ACKNOWLEDGEMENTS

First and foremost I would like to express my gratitude towards FCT – Fundação para a Ciencia e Tecnologia – which grant support this work's development. I would also like to thank my adviser, Carla Ferreira, whose consistent help was determinant for this work's success.

To my friends, Danna Krupka, André Rodrigues and Dymytry Krupka. Thank you for keeping me sane when all hell broke lose. To my friends on the other side of the globe, Maddalena Menabue and Matteo Doria, thank you for making my days a joy.

To my parents, which always make the impossible come true. This wouldn't be possible without your unconditional support.

Finally I would like to thank my brother for believing in me even when I didn't.

*If we knew what it was we were doing, it would not be called research, would it?*

# Abstract

Microservice architectures are an emergent technology that builds business logic into a suite of small services. Each microservice runs in its process and the communication is made through lightweight mechanisms, usually HTTP resource API. These architectures are built upon independently deployable and, supposedly, reliable pieces of software that may, or may not, have been developed by the team using it. Nowadays, industries are dangerously migrating into microservice architectures without an effective and automatic process for testing the software being used. Furthermore, current API specification languages are not expressive enough to be used for testing purposes. To solve this problem it is necessary to extend currently broadly used API specification languages. APOSTL is a specification language to annotate APIs' specifications based on first-order logic, with some restrictions. It has the purpose of extending the currently used API description languages with properties that can be useful for testing purposes, transforming these description documents into useful testing artifacts. Besides providing information needed for testing an application, APOSTL also provides an API with semantic. This additional information is then leveraged to automate microservice testing.

The work developed in this thesis aims to fully automate the microservice testing process. It is achieved by the implementation of PETIT a tool able to test microservices when provided with an OpenAPI Specification document, written in JSON and properly annotated with the previously proposed specification language, APOSTL.

The tool is able to analyze microservices independently from the source code availability.

# Resumo

As arquitecturas de microserviços são uma tecnologia emergente que constrói lógica empresarial através de um aglomerado de pequenos serviços, onde cada um deles corre num processo independente e a comunicação é feita a partir de mecanismos de comunicação leves, usualmente HTTP com APIs para recursos. Estas arquitecturas são construídas com base em software desenvolvido de forma independente, supostamente fiável, e que pode, ou não, ter sido desenvolvido pela mesma equipa que o utiliza. Actualmente, a indústria está a migrar, de forma perigosa, para arquitecturas de microserviços sem que exista um processo automatizado e eficiente para testar o software que estão a utilizar. Além disto, as linguagens de descrição de APIs actualmente utilizadas não são suficientemente expressivas para serem usadas para fins de teste. Para resolver este problema, é necessário extender as linguages de descrição de APIs mais utilizadas. APOSTL é uma linguagem de especificação para anotar descrições de APIs, baseada em lógica de primeira ordem. Tem como propósito extender linguagens de descrição de APIs com propriedades úteis para fins de teste, transformando os documentos de descrição em artefactos de teste úteis. Para além de fornecer informação útil para fins de teste, a APOSTL também dota a API com semântica. Esta informação adicional pode ser utilizada para automatizar o processo de teste de microserviços.

O trabalho desenvolvido nesta tese ambiciona automatizar totalmente o processo de teste de microserviços. Este objectivo é atingido com a implementação da PETIT, uma ferramenta capaz de testar microserviços apenas com a sua especificação, escrita em JSON, e devidamente anotada com fórmulas em APOSTL.

A ferramenta de teste desenvolvida é capaz de analisar microserviços independentemente da disponibilidade do código fonte.

**Palavras-chave:** teste automatizado, microserviços, testes de caixa-negra, desenho por contracto, geração de dados de teste

# Contents

# LIST OF FIGURES

# LIST OF TABLES

# LISTINGS

# Introduction

This chapter presents the context for the problem as well as the motivation to solve it. It also briefly describes the implemented solution, this work's contributions and a brief description of this document's structure.

## 1.1 Context

Microservice architectures are an emergent technology that builds business logic into a suite of small services, each running in its own process and communicating through lightweight mechanisms, usually HTTP resource API.

Microservice's code can be hidden to client applications which makes them black-box systems. In order to test such systems, one needs access to its specification. Current API specification languages have only information about the types, e.g., the operation responsible for adding a pet has in its specification information about what should be carried in the request – the representation of the new pet (name, photo, owner information) –, and information about the response contents, typically, an HTTP code according to the operation success or failure. This information is not enough to meaningfully and efficiently test microservices. In order to test such systems, it is necessary to know which properties should be guaranteed before and after an action call. Current API specification languages are not expressive enough to be able to provide these kind of properties – invariants, pre and postconditions. Thus, beyond the need for an efficient method to test microservices, there is the need for extending current API specification languages in order to be able to specify these logical conditions. In the previous example, one possible precondition could be that a request made to obtain a pet given its identifier should respond with the HTTP code 404 (not found); one possible postcondition could be that making a request to obtain a pet with the same inserted identifier should respond with the previously inserted pet object.

## 1.2 Motivation

Nowadays, industries are dangerously migrating into microservice architectures without an effective and automatic process for testing the software being used. Microservice architectures are built upon independently deployable and, supposedly, reliable pieces of software that may, or may not, have been developed by the team using it. How can one, effectively, test such services if the code is not accessible? The current practices of testing microservices consist of manually producing requests and checking the requests' responses and, therefore, are not reliable. Hence, the motivation behind this thesis lies on the fact that there is no trustworthy automatic process for testing microservices as a black-box.

The current way of specifying microservices' APIs are not suitable to testing, meaning APIs contain little to no information that aids in the microservice testing process. Thus, there is also a demand to develop an extension to current API specification languages in order to add useful information that can improve testing results.

This thesis problem can be approached in two different, equally useful, ways: the first, and more obvious, testing microservices as a black-box, not having access to its code; the second, verifying if a given microservice implementation diverges from its specification.

## 1.3 Proposed Solution

In this thesis it is proposed a new methodology for automatically testing microservices having only access to its API description. The developed tool, PETIT – aPi tEsTIngTool –, is able to test microservices when provided with an OpenAPI specification document, written in JSON, properly annotated with the proposed specification language, APOSTL – API PrOperty SpecificaTion Language. These annotations consist mainly, but not exclusively, of invariants, pre and postconditions written at the cost of the same API's operations.

Besides making requests to the API and evaluating the obtained results, PETIT is also able to generate the test data that is used to perform the tests and evaluate whether an API or an API operation is, in fact, according to its specification. As such, PETIT is composed by a parser – to parse the OpenAPI Specification document –, an input generator – responsible for all test data generations –, an APOSTL formula parser – to check whether an APOSTL formula is according to its grammar –, an HTTP manager component – responsible for managing all HTTP interactions between PETIT and the microservice being tested –, and, finally, the tester and evaluator component – which, as the name suggests, is responsible for the testing, so to speak, and for the formulas' evaluation.

In short, PETIT generates input, performs requests to the specified operations and, finally, evaluates the obtained results.

## 1.4 Contributions

This work contributions are an API specification language developed to specify API contracts, and an algorithm which automatically generates, meaningful, not redundant, test data to test microservices, based on its extended specification.

The specification language adds invariants, pre and postconditions to an already existing API description. The developed specification language lacks expressiveness when compared to others, e.g., HeadREST [1]. However, the fact that the specification is built from API pure operations makes it easier to use and understand. Using the operations from the API itself makes the specification closer to what programmers are used to write, thus, gaining in terms of usability.

A tool is developed to integrate the test case generation algorithm with the ability to automatically make requests to microservices, and check if the obtained response is verified by the oracle. The tool provides the user with the ability to test several APIs at once – as long as they are specified in the same document – to study the interactions between them. The operations are divided into three categories – constructors, observers, and mutators. The operation order within each category is selected randomly at the beginning of each execution. The user has the ability to control the order in which these categories are being tested, as well as the granularity of the output produced by the tool.

In short, the main contributions are an API description language, and a tool that fully automates the process of testing microservices, given a microservice specification.

## 1.5 Document Structure

The remaining of this document is organised as follows:

**Chapter 2 - Background** provides information on key concepts necessary to understand this work's development, more precisely, software testing techniques – white and black-box testing –, what are microservices and from what they evolved from, and an example of an API description language – OpenAPI Specification.

**Chapter 3 - Related Work** besides presenting some tools that automate software's testing process, this chapter also introduces relevant black-box testing techniques that can be applied to this thesis problem.

**Chapter 4 - Solution Design** describes the design process for both PETIT and APOSTL. It also illustrates how to use PETIT and APOSTL with an example – tournaments' application. This chapter also describes PETIT's architecture and all its possible outcomes.

**Chapter 5 - Solution Implementation** describes how PETIT and APOSTL are implemented. This chapter is compartmentalized in two sections, the first being responsible for APOSTL's implementation, and the second for PETIT's implementation. As such, the first section provides insight on how APOSTL is integrated with OpenAPI Specification, and a formal definition of APOSTL's grammar. The second, provides information on the testing methodology implemented by PETIT, and a description of all its architectural components.

**Chapter 6 - Evaluation** analyses PETIT's tests results when testing a correct implementation of the tournaments' application, as well as a faulty one. Implementation errors are incrementally added in order to ascertain if PETIT finds them and, if it does, how useful is its output.

**Chapter 7 - Conclusions and Future Work** provides this work's conclusions and presents what can be improved in both PETIT and APOSTL.

# Background

This chapter presents essential topics that aid in the comprehension of this thesis subject – invariant-driven automated testing applied to microservices. The first section describes program verification; next, there is a description of Hoare's logic, which is essential to understand program's specifications; it also explains what is design by contract, an approach to software design. Software testing section includes a brief introduction to different testing strategies: black-box and white-box testing. The following section aims to explain what are microservice architectures as well as service-oriented architectures, where both these concepts came from, their necessity and why microservices' popularity is rising. Hereupon, this section aims to explain what is software testing as well as what is, in this case, the software under test – microservices.

## 2.1 Program Verification

Being able to formally guarantee a program's correctness has been a constant problem during software development. To tackle this, it was necessary to develop some way of describing a program's expected behaviour: a program specification. Although this might seem a good idea, writing correct specifications is not easy and not always adopted by developers: besides having to write the program, they also have to reason about all possible correct program states and describe them. This results in incomplete specifications that might not match the written program nor guarantee its correctness.

To solve this problem the concept of *program analysis* arises. A program can be analysed statically or dynamically. If the analysis is *static*, it happens at compile time – based on the program's source code – meaning the program is not executed. This guarantees that if the program satisfies a property, then all its executions will satisfy that same property. Static analysis finds weaknesses in an early stage of development, resulting in less expensive fixes. If the program analysis happens to be *dynamic*, the program is executed against a set of test cases. It is extremely important to choose an adequate set of test cases: the test set should test as many different program states as possible. If test cases follow this rule, dynamic analysis can be considered more effective than static analysis.

Although both analysis approaches can be performed independently, the most effective way of analysing a program is to combine them: a static analysis should be performed followed by a dynamic analysis. On one hand, defects such as unreachable code, undeclared (or unused) variables, and uncalled functions are not detected in dynamic analysis. On the other hand, static analysis can produce false positives by, e.g., taking into account a condition that may never be true.

This thesis lies on dynamic program analysis, since its purpose is to automate microservice testing.

## 2.2  Hoare's Logic

Hoare's logic was first introduced by Hoare in 1969 [2] with the purpose of providing a logical basis for proofs of the properties of a program, e.g., the most important property of a program is whether it carries out its intended goal. This goal can be specified by making general assertions on the relevant variables' values, after the program's execution – rather than specifying particular values, assertions describe general value's properties and relationships between them.

Hoare also states that the validity of a program's outcome depends on the values taken by the variables before the program is initiated. This means one can also define assertions in the same way as the ones used to describe the results obtained upon termination. Hence, a new notation was introduced to connect precondition properties $P$, program execution $Q$ and properties describing the expected results $R$:

$$P \{Q\} R$$

This notation can be interpreted as "if the assertion $P$ is true before initiation of a program $Q$, then the assertion $R$ will be true on its completion" [2]. Assuming the absence of side effects on the evaluation of expressions and conditions, Hoare described the following axiom and rules:

1. Axiom of Assignment

   Considering the assignment $x := f$, if any assertion $P(x)$ is true after the assignment, it must also be true on the value of $f$ before the assignment, i.e., $P(f)$ must also be true before the assignment.

2. Rules of Consequence

   If the execution of a program $Q$ ensures the truth of assertion $R$, then it also ensures the truth of every assertion logically implied by $R$ [2]. Moreover, the same is applied to precondition properties: if $Q$'s execution ensures the truthiness of $P$, then it also ensures that every assertion logically equivalent to $P$ is true.

3. Rule of Composition

A program is a sequence of statements executed one after another. Thus, a program $Q$ can be defined as the sequence of all it's $n$ statements: $Q = (Q_1; Q_2; Q_3; ... ; Q_n)$. In formal terms, the rule of composition is:

$$\text{IF} \quad P\{Q_1\}R_1 \quad \text{AND} \quad R_1\{Q_2\}R$$
$$\text{THEN} \quad P\{(Q_1; Q_2)\}R$$

This means that if the resulting outcome of executing $Q_1$ satisfies $Q_2$'s precondition, and $Q_2$ satisfies the final outcome condition $R$, then the whole program $Q$ – sequence of $Q_1$ and $Q_2$ – will produce the intended result.

4. Rule of Iteration

Considering the program $Q = $ **while** $B$ **do** $S$, the rule of iteration can be defined as follows:

$$\text{IF} \quad P \text{ AND } B\{S\}P$$
$$\text{THEN} \quad P\{\textbf{while } B \textbf{ do } S\} \neg B \text{ AND } P$$

$P$ is a property that must be true on the loop's life cycle, i.e., before entering the loop, in all its iterations and on loop's completion. $B$ is the loop's entering condition, meaning that if $B$ holds, then $S$ is executed, otherwise the loop terminates. Thus, $B$ is assumed true upon initiation of the loop and false upon the loop's completion.

Although the described rules can be used to construct the proof of properties of simple programs, they are not sufficient to prove that a program terminates, e.g. as a result of an infinite loop. Hence, $P\{Q\}R$ should be interpreted as "provided that the program terminates, the properties of its results are described by $R$" [2].

## 2.3 Design by Contract

Design by contract, applied to object-oriented architectures, was first introduced by Meyer [3] with the goal of improving software *reliability*, which can be defined as the combination of correctness and robustness, i.e., the absence of bugs. The concept of reliable software is often associated with *defensive programming* techniques, where the programmer wraps its code with as many checks as possible, even if they are redundant. Although this technique may prevent some disasters, it can also cause new ones: introducing redundant code is never a good idea, either because it makes the code harder to understand, or because new bugs are directly introduced in the new checks. Thereby, guaranteeing software reliability requires a more systematic approach, thus, arising the notion of *design by contract*.

Inspired by the work on program proving and systematic program construction of Hoare [2], Floyd [4] and Dijkstra [5], Meyer created the notion of *contract* based on contracts performed in modern society where both parts, the contractor and the client, have obligations and benefits. Furthermore, an obligation for one of the parties is a benefit for the other. Applying this concept to software development is straightforward: if the execution of a task depends on a routine call to handle a subtask, the relationship between the client routine (the caller) and the called routine (the supplier) needs to be specified. These relationships are specified through assertions – predicates – that can be:

**Preconditions** are applied to individual routines. Preconditions describe the state in which the program must be before the call of a routine. If a precondition does not hold, the client code violated the contract, and the effect of the called routine is undefined and may, or may not, carry its intended purpose. If no precondition is specified – or the predicate is *true* –, all program states are accepted.

**Postconditions** are applied to individual routines. Postconditions describe the state of the program after the routine call. If a postcondition is violated, the supplier code has a bug, thus violating the contract. If no postcondition is specified, all program states are accepted after the routine's execution.

**Invariants** constraint all the routines of a class. Invariants are properties that must ever hold, in any circumstance. Hence, it must hold upon the creation of a class instance, and hold before and after every execution of every routine the class offers.

Assertions do not aim to specify special cases. Instead, they specify expected cases. Special cases should be handled through standard conditional control structures, e.g., if statements.

Pre and postcondition's "strength" should be carefully thought. While strong preconditions put a burden on the client side, weak ones are a burden in the supplier code. Choosing between the two is a matter of preference, though the key criterion should be to always minimize architecture's complexity.

## 2.4 Software Testing

According to Myers et al. [6], "testing is the process of executing a program with the intent of finding errors" and "an unsuccessful test case is one that causes a program to produce the correct result without finding any errors".

According to Fowler [30], software developers should write self-testing code, so that the testing process should be fully automated. Developers should create a test suite that can be automatically run against the code to be tested. The test suite should be built in such way that when all tests pass, one should be confident enough to release the software to production. Hereupon, there's a necessity of defining rigorous methodologies to automatically generate trustworthy test suites that can be also executed automatically.

Software testing can be compartmentalized in two main strategies: white-box testing and black-box testing. There are several methodologies that follow each strategy and wouldn't be realistic to approach all of them in this document. Thus, a few representative ones were chosen. Both strategies and methodologies are discussed in detail on the following subsections.

Complete test coverage is, generally, impossible to achieve. This affirmation is properly justified in the following sections.

### 2.4.1 White-Box Testing

White-box – or *logic-driven* – is a testing strategy where the software tester can go through the subject program's implementation. Therefore, the test cases are derived from the program's logic [7].

Hypothetically, achieving complete test coverage with a white-box testing strategy should be through *exhaustive path testing*, which derives a control flow graph from the implementation and then aims to build a test battery that executes all possible control flow paths. Although all the paths are covered, one cannot conclude the program is completely tested either because *exhaustive path testing* does not guarantee the program matches its specification, the program might have missing paths, and covering all paths does not check for *data-sensitive* errors.

Since the focus of this thesis is on automated testing of microservices from its specification, white-box testing techniques will not be further explored. More information on the subject can be found in the survey by Anand et al. [8].

### 2.4.2 Black-Box Testing

Black-box testing, also known as *input/output-driven testing* [7], is a testing strategy where the software tester is completely unaware of the program's implementation: its internal behaviour and structure are unknown. Instead, the tester will have to derive test data only from the program's specification.

Achieving complete test coverage using a black-box testing strategy implies that the program should be tested with not only all values in the input domain but also with all possible inputs. Testing following such criterion – *exhaustive input testing* – can produce an infinite number of test cases thus, becoming impossible to achieve in an acceptable time period.

In the following chapter some black-box testing techniques are introduced, since they're the ones applicable to this thesis subject.

## 2.5 Microservices

In order to explain why, nowadays, microservice architectures are preferred over service-oriented architectures, it is necessary to give a step back and understand why the need of a different architecture arose in the first place.

In this section there is a brief explanation on how these software paradigms emerged as well as definitions of their core components. Since both services and microservices are available through APIs, this section also features OpenAPI, a standard for API descriptions.

### 2.5.1 Service-Oriented Architecture

According to Shadija et al. [9], in a service-oriented architecture a service is an entity, accessible through an interface (API), encapsulating various components to provide an individual business function. Furthermore, a component can be a service if it's wrapped by a service layer.

The notion of component emerged when object-oriented architecture was not enough to fulfill the rising need of working at a higher level of granularity, i.e., having more functionality into a single, independently replaceable and upgradeable entity [31]. As such, component-based system development was the next big thing where systems were composed by components and these consisted of several objects enclosed together.

In a service-oriented architecture services are connected through a robust and heavy mechanism called Enterprise Service Bus (ESB) [9]. In spite of its robustness, this structure constraints the scalability of applications according to the business needs. For this reason, service-oriented architectures hamper the evolutionary design of applications and, once more, a need for a change of paradigm arises.

### 2.5.2 Microservice Architecture

Fowler [31] describes a microservice architecture as being the development of applications "as a suite of small services, each running in its own process and communicating with lightweight mechanisms, often an HTTP resource API". However, as the name suggests, shouldn't microservices be small portions of software? Not necessarily. According to Shadija et al. [9], the granularity of a microservice is an important part of the architecture. Furthermore, having fine grained microservices can introduce an overhead on managing the whole application. Hence, microservices are not necessarily small portions of software, as the name wrongly suggests.

The microservice architecture contrasts with more conservative forms of software development in the sense that a traditional application has all its functionality into one process and, as needed, it scales by replication into several servers. On the other hand, an application built according to a microservice architecture has its functionality spread into multiple services and it scales by replicating only the needed functionalities on a server [31].

The motivation behind the creation of microservices was mainly scalability. A microservice architecture specifies end points with the associated business logic [9]. Microservices and client applications communicate through Hyper-Text Transfer Protocol (HTTP) request-response via well specified endpoints on the microservice API. By using sophisticated endpoints, microservices are able to adapt to the needs of an ever-growing business logic. Since the application architecture is decentralized and the communication between microservices is cheap and easy, more logic can be implemented within microservices.

The microservice architecture aims to build decoupled and modular applications. Rather than using a complex communicating systems like an enterprise service bus, microservice developers prefer the approach "*smart end points and dumb pipes*", i.e., having a simpler middleware architecture and communicating through HTTP request-response with resource API's and lightweight messaging [31].

### 2.5.3 OpenAPI Specification

Representational State Transfer (REST) is an architectural style to develop web services. Its nuclear concept are *resources*. To identify resources involved in component interactions, REST uses a *resource identifier* [1]. Since resources can be accessed and modified concurrently through various components, a *resource representation* is used to capture the current, or intended, state of that resource. Those representations are then transferred between components through REST interactions. REST systems communicate over HTTP and are made available to other systems as web resources identified by URIs [1]. Since the communication is through HTTP, the interactions are all HTTP verbs: GET, POST, PUT and DELETE to retrieve, add, update or remove resources. Additional information can be sent in the headers and the body of an HTTP request, and the results always include a response as well as a response status code.

RESTful systems are the ones developed using the REST architecture. These systems are an agglomerate of resources and their respective actions. A RESTful API is a set of resource identifiers as well as all the actions that can be performed on each resource.

OpenAPI Specification (OAS), formerly Swagger Specification [32], was created with the purpose of standardizing the way RESTful web services are described. OpenAPI is a description format for services' APIs that is language independent, portable and open [33]. Figure 2.1 contains an OpenAPI description of a pet store's pet management system found in [34]. It shows four actions that can be performed, their URI and a textual description.

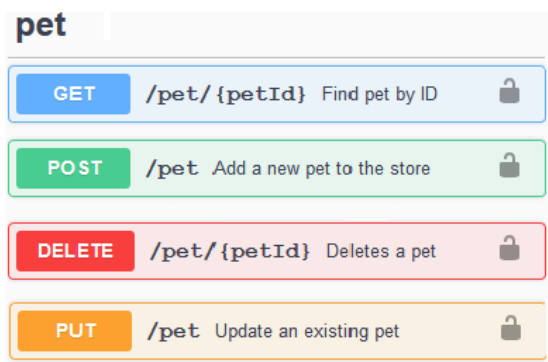

Figure 2.1: Pet store API example.

Figure 2.2 shows all information OAS provides for each operation. In this example, operation POST in the URL "*/pet*" expects to receive a JavaScript object – representing a pet – as parameter, and returns the HTTP code 405 in case of receiving an invalid input.

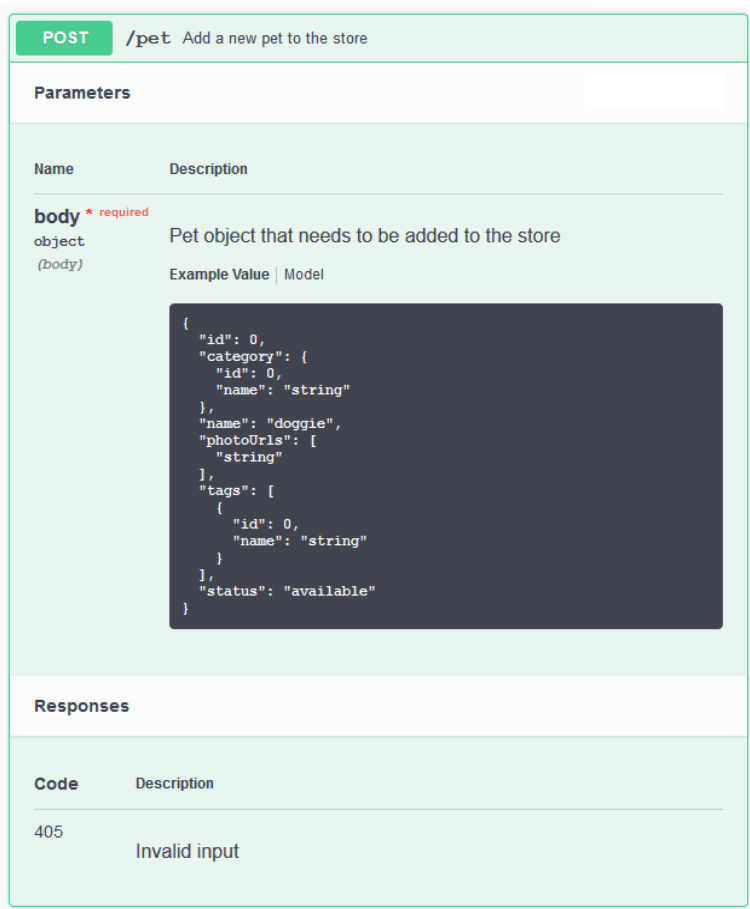

Figure 2.2: Operation POST expanded.

Although OAS files can be written in JSON or YAML, all examples will be presented in YAML for readability purposes. An OpenAPI specification file has the following structure [35]:

**Information** 2.1 contains the API's current version, its title and all applicable licenses.

```
1  info:
2    version: 1.0.0
3    title: Swagger Petstore
4    license:
5      name: MIT
```

Listing 2.1: YAML object for the API information description.

**Servers** 2.2 have information on all API servers and their URLs. Different servers can be used to implement an API, e.g. a sandbox server can be used with test data.

```
1  servers:
2    - url: http://petstore.swagger.io/v1
```

Listing 2.2: YAML object for the API servers.

**Paths** 2.3 defines API endpoints. Each endpoint is comprised of all HTTP methods it supports. Since each endpoint can be associated with different operations, the definition of each operation is achieved by using a *Path Item* object which, in turn, and depending on the HTTP method, has the summary, parameters array, request body, and the responses array.

```
1  paths:
2    /pets/{petId}:
3      get:
4        summary: Info for a specific pet
5        parameters:
6          - name: petId
7            in: path
8            required: true
9            description: The id of the pet to retrieve
10           schema:
11             type: string
12       responses:
13         '200':
14           description: Expected response to a valid request
15           content:
16             application/json:
17               schema:
18                 $ref: "#/components/schemas/Pet"
19         default:
20           description: unexpected error
21           content:
```

```
22                    application/json:
23                       schema:
24                          $ref: "#/components/schemas/Error"
```

Listing 2.3: YAML object for the API servers.

**Components** 2.4 to condense the file size and avoid information repetition, the components section is where the data structures used throughout the API are defined. Within components schemas can be defined. A schema has a type an array of properties and an array indicating the required properties. Schemas are referenced throughout the OAS document using the keyword *$ref*.

```
1   components:
2     schemas:
3       Pet:
4         type: object
5         required:
6           - id
7           - name
8         properties:
9           id:
10             type: integer
11             format: int64
12           name:
13             type: string
14           tag:
15             type: string
```

Listing 2.4: YAML object for the API servers

OAS does not have any information on the state of the system prior nor post operation execution. However, it supports the addition of custom properties. By using this mechanism, it is possible to extend OAS in order to add information about the valid states in which the system will perform as expected, as well as all information required to generate valid testing data. Hence, the addition of new properties, i.e. extending OAS, can be achieved by prefixing the new property with "x-".

All APOSTL annotations take advantage of OAS's ability to add custom properties. These annotations are enclosed only within the following properties:

**x-invariants** can be found in the beginning of an API description and contains a list of all API's invariants.

**x-requires** can be found in the beginning of an operation description and contains a list of all operation's preconditions.

**x-ensures** can be found in the beginning of an operation description, after the *x-requires* property, and contains a list of all operation's postconditions.

**x-regex** can be found either within the description of a model's property or in the description of an operation parameter and contains a regular expression that correctly generates the property or parameter.

## Related Work

This chapter presents some black-box testing techniques as well as a comparison between them. It also features some tools that automatically generate test data in different circumstances. Since the purpose of this thesis is to, ultimately, fully automate the testing process of microservices, the presented tools are intrinsically related to this subject. A brief description of HeadREST – a more expressive specification language than the ones currently used in the industry – can also be found in this chapter. There are also described some industry's current practices concerning microservice testing.

## 3.1 Black-Box Testing Techniques

### 3.1.1 Random Testing

Random testing is one of the most popular black-box testing methods [8]. Its implementation is not complex and when the system's specification is incomplete it is the only applicable testing technique.

An operational profile can be obtained through partitioning the input domain and assigning a probability to each partition. For programs where the operational profile is known, for whose domain a pseudorandom number generator is available, and for which there is an effective oracle, the general idea behind random testing follows the steps [10]:

1. Selection of a test case size, $N$.

2. Assign a probability $p_i$ to each one of the $K$ operational's profile partitions. Each partition has an unique domain, hence partition $i$ is now mentioned as $D_i$.

3. Generation of $N_i$ test cases – from the pseudorandom number generator – for partition $D_i$ such that $N_i = p_i N$, for $1 \leq i \leq K$, i.e., the generator will pick a number within $D_i$ with probability $p_i$. All these $N_i$ form the test set.

4. Execute the program with the generated inputs.

5. Use the oracle – function that checks if a result satisfies the system's requirements – to detect any failures. If any failures are detected the software suffers adjustments and is, once more, tested with a new pseudorandom test set with the same size. When no failures are detected for a test set with size $N$, the testing is complete.

For programs where inputs are not straightforward – e.g. objects instead of only numbers and strings –, partitions are defined for sequences of inputs, i.e., the operational profile describes "classes of input sequences" [10] and the previously described procedure can be used to randomly select a test set of sequences. The most common case is random testing being applied with only a requirements document that has no information about input sequences by the absence of usage information. Thus, it is common that the operational profile is not available since the input is not made up of single values. When this happens, random testing is applied with a uniform distribution, i.e., attributing the same selection probability for every class of input sequences.

### 3.1.2 Specification-Based Testing

The foundation of every specification-based testing technique are user requirements – generally specified in a formal logical language – regarding the software's functional behaviour. By having the requirements formally expressed, it is possible to automate both test case generation and verdict construction. The general steps of specification-based testing are the following [11]:

1. Test Case Generation:

   Generation of a test case $i$ in which the preconditions present in the user requirements are satisfied.

2. Test Case Execution:

   Execution of test case $i$ on the system under test produces a result $o$.

3. Oracle:

   Analysis of the pair ($i$, $o$) with the requirements through a constraint checker to determine a verdict about the generated test case $i$. If the pair satisfies the requirements the test case $i$ passes, otherwise it fails.

### 3.1.3 Learning-Based Testing

Learning-Based testing emerged with the purpose of improving specification-based black-box testing. This is achieved by the automatic generation of a vast number of test cases within a reasonable time frame and, at the same time, improving test case quality by taking into account the result of previously executed test cases.

In LBT all learning can be classified as *active learning* [11] since different algorithms are used to generate new queries (test cases) during the learning process. Three types of queries can be identified [11]:

**Model checking queries**  generated by model checkers

**Structural queries**  generated by learning algorithms

**Random queries**  generated by random data generators

Test efficiency – here defined as the number of queries needed to find an error – is influenced by query type. Therefore, queries should be seen as "expensive", meaning the most efficient type of query should be chosen at all times. Empirical evidence shows that random queries result in the least efficient test cases [11]. Hence, LBT is an improvement to the pure random testing technique – unless the error distribution of the system under testing is very large –, since it finds errors that would be hard to find by using random testing, in a more time-efficient manner.

The novelty of learning-based testing, against the previously described process of specification-based testing, is the introduction of a *feedback loop* [11] into the process previously described, which can be accomplished by introducing a learning algorithm with the purpose of trying to infer a model of the system based on the already generated test data, i.e, pairs ($i$, $o$). This model is then automatically analysed with the intent of finding counterexamples in the learned model to the requirements' correctness, i.e. to check if the learned model diverges from the specification. The newly found counterexamples are then treated as a new test case. If the model is accurate then there's a high probability that the new test case will incur in an error – expected result different from the obtained result. The accuracy of the model tends to improve over time since it is constantly fed with new, already executed, test cases.

The choice of a learning algorithm should not be taken lightly since it infers the models used to generate new test data. Further information regarding suitable learning-based testing algorithms can be found in the following articles by Meinke [12], Meinke and Sindhu [13].

### 3.1.4 Adaptive Random Testing

Adaptive Random Testing (ART) was first introduced by Chen et al. [14] and it was developed to improve the failure-detection effectiveness of random resting. It relies on "empirical observations showing that many program faults result in failures in contiguous areas of the input domain" [14]. Hence, one can infer that regions of the input domain where the software produces results according to the specification, i.e., are correct, are also contiguous. Therefore, if a set of previously executed test cases have not lead to failures, the likelihood that test cases farther away from the previously executed ones will lead to a failure increase. Therefore, if previous tests have not led to failures, new test cases should be distant from the already executed ones.

Since the objective of a software tester is to maximize the number of detected faults and these faults are proven to occur in contiguous regions of the input domain, there's a need to change the pure random testing technique in some way that introduces some diversity into the generated test cases, i.e., test cases should be evenly spread through the input domain.

In order to implement the ART technique, one can follow several approaches. The even spread of test cases can be achieved from different algorithms following each approach. The most commonly used approaches are the following [8]:

**Selection of the best test case from a set of test cases:** This technique starts by computing a set of random inputs where the best candidate should be drawn. The most commonly used algorithm implementing this approach is *Fixed Size Candidate Set ART* (FSCS-ART) [15]. Since this was the first algorithm implementing ART and, according to [8], has been the most cited ART algorithm, it is the one chosen to illustrate the technique in this document.

### *Fixed-Size-Candidate-Set Adaptive Random Testing Algorithm*

Whenever a new test case has to be chosen, a fixed-size candidate set of random inputs is generated. For each candidate set a selection criteria is applied to select the best candidate as the next test case. The selection criteria can be, amongst others, maxi-min or maxi-sum. It is necessary to compute the distance – or some measure of dissimilarity, for non-numerical inputs – between the previously executed test case and all the candidates. If the selection criteria is maxi-min then the candidate farther away from the previously executed test case is the chosen one. If the selection criteria is maxi-sum, the distances between each candidate and all the previous executed test cases are added together being the candidate with the greater sum value the chosen one.

One of the problems with these algorithms is that a distance – or dissimilarity – measure is not naturally defined for non-numerical inputs.

**Exclusion:** All methods following the Exclusion approach have an exclusion region for each previously executed test case. Random inputs are generated until one input is outside all exclusion regions. When an input following this criteria is generated, it is selected as the next test case to be executed and, consequently, an exclusion region is defined around it.

**Partitioning:** The Partitioning approach demands the input domain to be divided into several partitions. The next partition from where the next test case is generated is chosen by taking into account the previously executed test cases, i.e., from where they were drawn. Further information on this subject can be found in the article by Chen et al. [15].

**Test Profiles:** In this approach, an unique test profile is developed in order to fulfill the requirement of even spreading of test cases throughout the input domain as opposed to random testing where the test profile commonly follows an uniform distribution. More information on test profiles can be found in the article by Liu et al. [16].

**Metric-Driven:** This approach has the peculiarity of using distribution metrics, such as discrepancy or dispersion, as selection criteria to the next test case to be executed. The usage of metrics as criteria has the purpose of evenly distribute test cases throughout the input domain.

Further information on different implementations of ART algorithms can be found in the following documents: Chen et al. [17, 18], Ciupa et al. [19], Lin et al. [20], Mayer [21], Shahbazi et al. [22] and Tappenden and Miller [23].

### 3.1.5 Discussion

Although all previously presented techniques can be applied to automatically generate test data for microservice testing, some are more suitable than others. A pure random approach is inadvisable, since it can produce redundant and meaningless data.

On the other hand, a learning-based testing technique can be used, since it is able to find errors typically hard to find with pure random testing. With the proper learning algorithm, the inferred system's model can be accurate enough for the tester to be able to affirm that the next generated test case will incur in an error.

Adaptive Random Testing technique, like LBT, is a major improvement to pure random testing. By assuming that faults result in failures in contiguous areas of the input domain, several approaches were developed to fulfill the requirement of test data being evenly spread throughout the input domain. Since this idea can incur in an undesirable overhead, it is necessary to choose the best ART approach as well as the best algorithm implementing it.

## 3.2 Tools for Automated Testing

Although these tools do not aim to test microservices directly, the process can be applicable to microservice testing.

### 3.2.1 QuickCheck

QuickCheck [24] is a tool that generates random test data for Haskell programs. Haskell is a purely functional programming language which makes programs written in it very well suited for automatic testing. This happens because pure functions, i.e., non side-effecting functions, are easier to test than side-effecting ones. Hence, small code portions can be tested separately, allowing the software tester to perform meticulous testing at a small granularity.

The authors state that a testing tool must be able to:

1. Determine whether a test has passed or failed:

   The user defines expected properties of the functions under test in a domain-specific language, designed by the authors.

2. Automatically generate suitable test cases:

   The technique used to generate test cases is random testing. Although it may seem a naive approach, the authors based their choice on results presented by Duran and Ntafos [25] showing that the difference in effectiveness of random testing and partition testing is small.

   Furthermore, it was a requirement that QuickCheck was a lightweight tool. Using more systematic methods (e.g. partition testing) would violate this requirement because some *adequacy test criteria* [24] needed to be reinterpreted before it could be applied to functional programs. Not to mention that applying these methods would require compiler modifications and hence bond QuickCheck to a particular implementation of Haskell, making their choice of using random testing very clear.

Since random testing is used, it is necessary to discuss the distribution of the test data. As stated above, the efficiency of random testing is maximized when the distribution of the test data is the same of the actual data. QuickCheck does not infer a distribution. Instead, the authors defined a *test data generation language*, allowing the tester to program a suitable generator, controlling the distribution of test cases.

### 3.2.2 JET

JET is an evolutionary testing tool [26] developed with the purpose of automating random testing of Java programs to detect as many inconsistencies as possible between the specification – written in Java Modeling Language (JML) – and its implementation. JET automatically generates test data – through a pure random approach –, executes the tests and determines the tests results – using a runtime assertion checker as an oracle –, thus fully automating the testing process.

Notwithstanding the utility of the tool by itself, there is an extension to JET, developed by Cheon and Rubio-Medrano [27], in which test data generation is not purely random. To randomly construct a Java object without having direct access to its internal state means the object has to be constructed via method calls. Thus, test data consists of sequences of method calls. Objects' methods are divided into three categories: constructors, mutators and observers. By using a pure random technique, method calls – constructors and mutators since observers do not contribute to objects' state alteration – are randomly selected, all at once, hence not ensuring the produced object is in a consistent state. A study shows that more than 50% of randomly generated test data are redundant [27]. Hereupon, the extensions' goal is to generate meaningful, not redundant, test data. This is achieved by constructing the object incrementally – i.e. not determining the call sequence at once –, ensuring the validity of each randomly selected method call. Hence, an object is constructed only by feasible method calls – verified by JML's assertion checker – guaranteeing the "randomly" generated object is in a consistent state. In order to solve the redundancy problem, when generating a new object, a pool of previously generated (and consistent) objects is used: an object is picked from the pool and then a new call sequence is appended to it, thus generating a new, consistent and not redundant object.

By using this approach, there is a minimum increase of 10% [27] in the number of successfully generated test cases.

### 3.2.3 Korat

Korat is a framework that uses specification-based testing to automate the testing process of Java programs [28]. Given a method's formal specification written in any specification language – as long as it can be translated to Java predicates –, Korat uses the precondition to generate test cases up to a given size. It then invokes the method on each generated test case and uses the post-condition as the oracle.

The most interesting aspect of Korat is the technique for test case generation: given a predicate and a bound on the size of its inputs, Korat generates all non-isomorphic inputs that verify the predicate, i.e., for which it returns true. In order to generate valid test cases for a method, Korat creates a class whose fields are the method's parameters, including the implicit parameter *this*. This class also has a predicate – function returning a *Boolean* value –, which is, essentially, the method's precondition. It then generates all distinct inputs for which the predicate returns true. Since the predicate is the method's precondition, all generated inputs are valid inputs.

To check the correctness of a method, all method's valid inputs are generated. Next, the method is invoked on each generated input, testing, in each iteration, if the produced output is correct, using the oracle. If it's not, then the input is a counterexample and the method under test is incorrect [28].

One of the most relevant experimental results using Korat is that theses results prove the feasibility of automatic test case generation for Java predicates even when the search space for inputs is very large [28].

### 3.2.4 Discussion

QuickCheck was developed with the purpose of randomly generating test data for functional programs. It uses a pure random testing strategy and does not even try to infer test data distribution. For these reasons, QuickCheck approach is considered to be the least valuable for the purpose of automatically generate test data in order to test microservices.

On the other hand, the extension to JET does not follow a pure random testing approach: test data is built incrementally and its validity verified in each iteration, leading to automatically generated, not redundant, test data. This approach can be, with some adaptations, applied to microservices: constructor methods can be POST actions, mutators can be PUT and DELETE actions and, observers can be GET actions. Hence, this technique can be used, with a few tweaks, to automatically generate test data for microservice testing.

The main idea behind Korat's is that by having both pre and postconditions, being able to automatically generate test cases based on the precondition – only generating valid test cases – and test the method's performance with the postcondition – the oracle. This approach can also be directly applied on microservice testing since pre and postconditions are assumed to be available. If the postcondition is not available, the oracle can be an invariant.

In short, both QuickCheck, the JET extension and Korat approaches can be used to test microservices, being the least preferable the pure random testing technique used by QuickCheck since it tends to produce an undesirable amount of meaningless data.

## 3.3 Extending OpenAPI: HeadREST

HeadREST is a language to describe RESTful APIs developed by Vasconcelos et al. as a part of Confident, a research project on the formal description of RESTful web services using type technology [1]. HeadREST allows to specify data properties and to observe server state changes through assertions. These assertions are Hoare triples of the form

$$\{\phi\} \ (a \ t) \ \{\psi\}$$

where $a \in \{\text{GET, POST, PUT, DELETE}\}$, $t$ is an URI – e.g., in figure 2.1, **/pet/{id}** – and both $\phi$ (precondition) and $\psi$ (postcondition) are predicates. This assertion should be interpreted as: if a request to execute action $a$ over the URI $t$ has data satisfying $\phi$ and $a$ is executed on a state satisfying $\phi$, then both the data carried by the response and the resulting state satisfy $\psi$ [1].

The motivation behind the creation of HeadREST lies on the fact that the current way of specifying APIs is mainly focused on the structure of the exchanged data and therefore, ignore the ability to relate different parts of the same data, the relationship between input and the service's state, and, finally, the relationship between input and output. Recalling the Pet Store example, figure 2.1: supposing a pet has an *owner* and this *owner* has a *name* and a *nickname*, there is no way, in the currently available API specification languages – e.g., OpenAPI Specification –, to specify that, e.g., the nickname must not have more than 15 characters. HeadREST is a more expressive way of specifying APIs, relying on two main ideas [1]:

- **Types** that allow to express data exchanged in the interactions and properties of server states

- **Pre and postconditions** to express the relationship between the input – what was sent in the request – and the output – what comes in the response.

To make OpenAPI suitable to be used for test case generation, a similar approach to HeadREST will be used.

## 3.4 Current Industrial Practices

Industry's most used tools to test microservices are described in this section with the purpose of illustrating the demand for a method/technique to fully automate the process of testing microservices.

### 3.4.1 Manual Testing

None of the following tools can be considered automated testing since test data is produced manually, the microservice is manually invoked once for each test, and the verification is not made by an oracle.

**cURL**  cURL, or client URL [36], is a project providing a library and a command-line tool to ease data retrieval through several protocols. When the chosen protocol is HTTP, the user is expected to provide the URL, the headers, and body of the request. In spite of the ultimate goal of this tool being data retrieval, is has been used to test microservices manually: the tester makes a request using cURL and then checks if the response matches the expectations. Needless to say this process is very time consuming and, therefore, not suitable to testing microservices in a large scale.

**Postman**  Postman's main goal [37] is to design, build and test APIs. However, it can also be used to test microservices by making requests, just like the previous tool, and comparing the obtained results with the expected ones. Postman can be used to manually test a microservice in the same way as cURL, with the only difference being that Postman provides an easy to use GUI. Postman also organizes requests in collections allowing the tester to reuse a previously done request.

### 3.4.2 Semi-Automated Testing

The following tools can be considered semi-automatic since results' validation is made automatically although test data needs to be provided by the tester.

**Dredd**  Dredd's main goal is to test API's implementations. Given the API's description document – supported languages are API Blueprint and Swagger [38] –, Dredd creates expectations based on requests and responses specified in the given document, then it requests resources to the API being tested, and verifies if the obtained results are according to the specification. For operations requiring parameters, Dredd uses values provided in the specification or, if none is present, Dredd generates some dummy values according to the provided schema (or data model) – e.g. Swagger's schema is defined in JSON [39]. In spite of Dredd being able to generate test data, it does not mean the generated data is valuable, i.e., it may not happen on a real situation. For this reason, Dredd is only a reliable testing tool if test data is provided by the tester.

**Postman** Postman eases manual testing, as seen previously, however, it has more interesting features: it also provides a way to kind of automate the testing process by allowing the tester to write scripts [40], in JavaScript, that are able to validate the obtained response.

# SOLUTION DESIGN

Microservices are commonly used as black-box systems, meaning its consumers are oblivious of its implementation. However, microservices are accompanied with APIs that can be used as test artifacts. Although these APIs are usually well documented, they lack essential information for testing purposes. As such, microservice's APIs need to be extended in order to accommodate contractual information (described in section 2.3) about each operation – pre and postconditions – and about the APIs' valid state – invariants. These additional annotations are written in APOSTL, a specification language for describing API invariants and operations' pre and postconditions. Microservices' APIs also have information about the data structures exchanged in each operation. Therefore, this data schema can be improved by including information on how each element can be generated. In short, having a microservice description document with information regarding the system's state prior and post an operation, and information regarding how a data structure can be generated provides us with all the information needed to automate the microservice testing process.

PETIT is an automated microservice testing tool which only requires the microservice specification properly annotated with APOSTL. This specification language has the particularity that all operations used to describe predicates need to be pure, meaning they cannot produce any side-effects to the microservice's state.

Figure 4.1 illustrates all the steps a user needs to perform in order to use PETIT. As shown in the figure, the user must first annotate the OAS file with its contract. The next step is to annotate the same file with the regular expressions, needed for the data generation. Once the OAS is complete, the user is ready to execute PETIT. Hence, one must specify the OAS document path and define the order in which operations' categories will be tested. Then, and optionally, one can specify the API testing order – random or sequential, the later meaning "the order as defined in the OAS document" – as well as the output form – verbose or standard mode. The standard execution only displays the testing results. If PETIT is executed in verbose mode the response contents of each operation will be shown. In the verbose mode execution there is also the need to specify the maximum number of REST resources to be displayed.

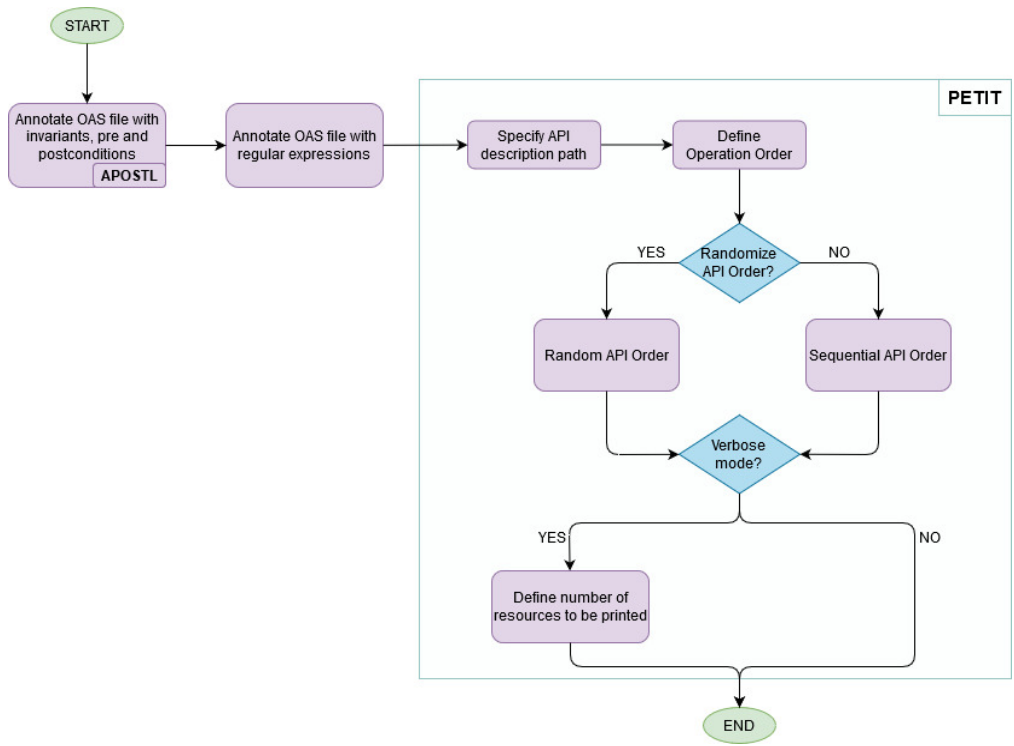

Figure 4.1: Steps needed to execute PETIT.

The testing methodology followed by PETIT begins with categorizing all APIs' operations into three disjoint sets: *mutators* composed by PUT and DELETE methods, *constructors* composed by POST methods, and *observers* composed by GET methods. This compartmentalization serves the purpose of manipulating the order in which each category is being tested. The operation order within each category is randomized.

The testing process of each API operation starts by checking if all API's invariants hold and, if they do, the testing process proceeds by generating or recycling the needed data, when applicable. Then, precondition verification begins and, if all conditions hold, the HTTP request is performed. Once a response is received, the postcondition verification takes place and the testing process is complete.

| Precondition | Request | Outcome |
|:---:|:---:|:---:|
| True | 200 | OK |
| True | 4XX | Failed (analyse execution trace) |
| False | 200 | NOT OK |
| False | 4XX | Failed (as expected) |

Table 4.1:  Operation test outcomes.

The possible test outcomes for a single operation are described in table 4.1. According to the outcomes presented in the table, when all preconditions hold (true) and the operation's response was not successful (4XX) the test failed, and there is the need to analyse the execution trace, e.g, this scenario usually happens when one is trying to retrieve a resource that was previously deleted. When the there is at least one precondition that does not hold (false) and the operation's response was not successful (4XX), the test has failed as expected, since the preconditions did not hold in the first place.

This chapter describes the design process behind both PETIT and APOSTL, as well as illustrate the fundamental concepts with an example application.

## 4.1 Tournaments' Application

In order to better understand how to use PETIT, consider a tournaments' application composed by two APIs – *players* and *tournaments* API. This application's purpose is to manage player's enrollments in different tournaments. As such, a player can be both enrolled and disenrolled from a tournament, as long as the number of enrolled players has not reached the tournament's capacity. Figures 4.4 and 4.5, respectively, depict player's and tournament's APIs.

The players API manages all player resources which are identified by the *playerNIF* property, and composed by the properties shown in figure 4.2. The property *tournaments* is a collection of the tournaments in which the player is enrolled. When expanded, it shows the tournament's schema, depicted in figure 4.3.

```
Player ∨ {
    address          string
                     example: The cupboard under the stairs, 4 Privet Drive, Little Whinging, Surrey
    email*           string
                     example: harry.potter@wizardmail.uk
    firstName*       string
                     example: Harry
    lastName         string
                     example: Potter
    phone            string
                     example: 999999999
    playerNIF        string
    tournaments
                     > [...]
}
```

Figure 4.2: Player schema from tournaments' application.

On the other hand, tournaments API manages all tournament resources which are identified by the *tournamentId* property and composed by the properties shown in figure 4.3. The property *players* is a collection of the players enrolled in the tournament. When expanded, it shows the player's schema, depicted in figure 4.2.

As seen in figure 4.4, player's API describes all operations responsible for managing a player resource. These operations are responsible for inserting, updating, retrieving and deleting a player from the system as well as retrieving a player's enrollments.

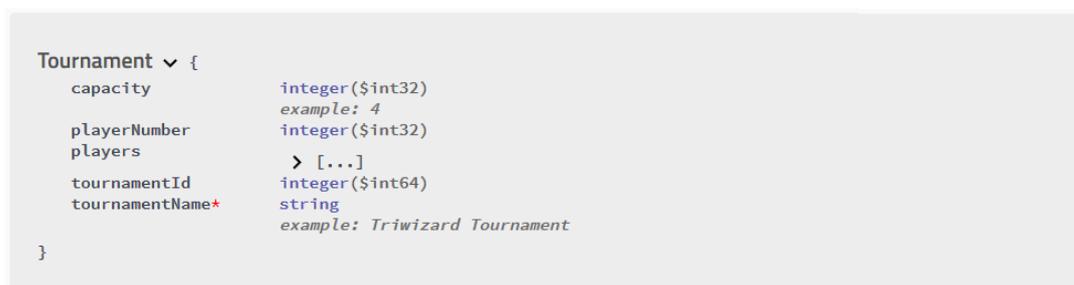

Figure 4.3: Tournament schema from tournaments' application.

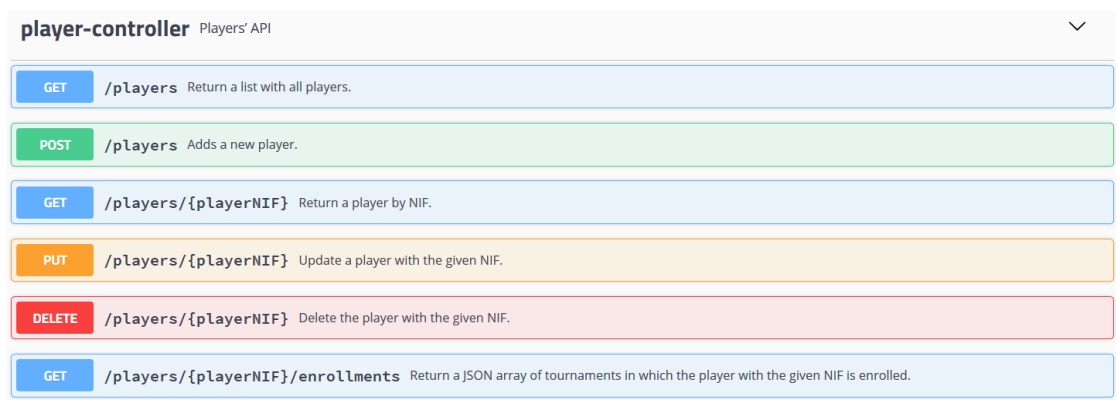

Figure 4.4: Player's API operations.

Similarly, the tournament's API, as seen in figure 4.5, describes operations responsible for managing a tournament resource and, as such, one can insert, update, retrieve, and delete a tournament, retrieve a tournament's capacity and its enrollments, as well as both enroll and disenroll a player from a tournament. Both APIs have operations to retrieve all their managed resources.

The tournaments' application is the case study used throughout this thesis and, as such, it will be frequently referenced in future chapters, serving as a base to explain the fundamental concepts both for the conditions written in APOSTL as well as the testing methodology implemented by PETIT.

## 4.2    Specification Language: APOSTL

APOSTL is a specification language to annotate APIs' specifications based on first-order logic. It has the purpose of extending the currently used API specification languages with properties that can be useful for testing purposes, transforming these documents into useful testing artifacts. Besides providing information needed for testing an application, APOSTL also provides an API with semantic, i.e., with these annotations one can easily understand each operation's logic.

APOSTL's main feature is the ability of writing logical conditions based on pure (without side-effects) API operations. These conditions are used to write operation contracts.

Figure 4.5: Tournament's API operations.

In the same way, APOSTL is also used to write API invariants. Although being initially designed for extending OAS, APOSTL can also be used with any API specification language that has the ability to be extended.

While developing APOSTL, there was a concern that was always present: usability. The problem with many specification languages is that in order to use them effectively, one needs to conquer a challenging learning curve. With APOSTL, the specification developer will only need to know a few intuitive keywords, basic knowledge of first order logic and its own API.

Considering the proposed example – the tournaments' application – and focusing on the operation responsible for inserting a player from players' API, one can derive some logical properties that should constitute this operation's contract:

**Precondition** Only a player that does not exist can be inserted.

**Postcondition** After the insertion, the player must be in the system.

This contract states that if the client follows the precondition then the server will ensure the postcondition is held. In APOSTL, these two conditions should be written only at the cost of pure operations which, in RESTful APIs, translates into GET operations. As such, one way of writing the contract for this operation is depicted in listing 4.1.

```
// Precondition
response_code(GET /players/{playerNIF}) == 404

// Postcondition
response_code(GET /players/{playerNIF}) == 200
response_body(this) == request_body(this)
```

Listing 4.1: Player's API POST player operation contract.

APOSTL takes advantage of the standardized HTTP codes. As seen in listing 4.1, the precondition states the response code of a request to get the player yet to be inserted must return the code 404 (resource not found). Similarly, the postcondition states that after the insertion, the same request should return the response code 200 (OK), meaning the player is persisted in the system. The second postcondition might not be as trivial as the previous one: the response body of the POST request must be equal to the same request's body. This condition ensures that what is returned form the server is exactly what was sent by the client.

With APOSTL one can also access the previous state of an API. The operation responsible for deleting a player makes use of this feature. This operation's contract is described in listing 4.2.

```
// Precondition
response_code(GET /players/{playerNIF}) == 200

// Postcondition
response_code(GET /players/{playerNIF}) == 404
response_body(this) == previous(response_body(GET /players/{playerNIF}))
```

Listing 4.2: Player's API DELETE player operation contract.

The precondition states that for a player to be deleted it must exist. The first postcondition states that, if the precondition holds, then the player is deleted from the system. The last postcondition, once again, is regarding the contents of the server's response: the response body must be equal to the response body from a request retrieving the same player **before** the current request is performed, i.e. the deletion.

APOSTL also allows the usage of quantifiers. For instance, one invariant for the *tournaments* API is depicted in listing 4.3.

```
// Invariant
for t in response_body(GET /tournaments) :-
response_body(GET /tournaments/{t.tournamentId}/enrollments).length <=
response_body(GET /tournaments/{t.tournamentId}/capacity)
```

Listing 4.3: Tournament's API invariant.

This invariant states that, for all tournament resources, the number of the tournament's enrolled players needs to be less or equal to the tournament's capacity.

### 4.2.1 Data Generation

Once all API operations are properly annotated with invariants, pre and postconditions, one can also provide information on how to generate exchanged data. This information is specified using regular expressions. Returning to the previous example – the tournaments' application –, and considering the operation responsible for retrieving a single player, partially specified in 6.8. This operation has a potentially interesting parameter, of the type string, *playerNIF*. The parameter schema of a regular OAS would normally just have the property *type*. However, an additional property was added, *x-regex*. If this property is present, PETIT will generate data according to the information described in the regular expression.

```yaml
"/players/{playerNIF}":
  get:
    summary: Return a player by NIF.
    x-requires:
    - T
    x-ensures:
    - T
    parameters:
    - name: playerNIF
      required: true
      schema:
        type: string
        x-regex: "(1|2)[0-9]{8}"
```

Listing 4.4: YAML object for Player's API get player operation.

As previously mention, APOSTL is based on first-order logic with some restrictions. The restrictions are mainly focused on nested conditions, e.g., APOSTL does not allow nested quantifiers nor quantifiers with more than one variable. Restrictions will be further discussed in the implementation chapter.

## 4.3 Testing Tool: PETIT

This thesis proposes a new methodology for automatically testing microservices, having only access to its API description file. The developed tool, PETIT, is able to test microservices when provided with an OAS document, written in JSON and properly annotated with the previously proposed specification language, APOSTL.

PETIT is made up of several components, each one being responsible for a different stage of the testing process. Its architecture, depicted in figure 4.6, shows not only the different components of PETIT, but also its execution flow, from the point where the specification file is provided to the API testing results.

As seen in figure 4.6, the OAS file is processed by the *specification parser* component, which is responsible for taking the information of the API description and make it available as Java objects. Thus, the *specification parser* produces a specification object and several schema objects. The schemas are used by the *input generator* component in order to only generate valid test data, i.e., valid JSON elements. The specification, in turn, is used by the *formula parser* which is responsible for not only replace the parameters with the generated test data, but also to analyse if the resulting formula is according to APOSTL. Finally, the *tester and evaluator* will, as the name implies, be responsible for testing the application and evaluating the results. As such, it verifies the invariants and preconditions and forwards the requests to the *HTTP manager* component, which has the purpose of performing all needed requests to the microservice, process and forward the received responses to the *tester and evaluator*. The *tester and evaluator* then evaluates the preconditions and invariants and outputs the API testing results.

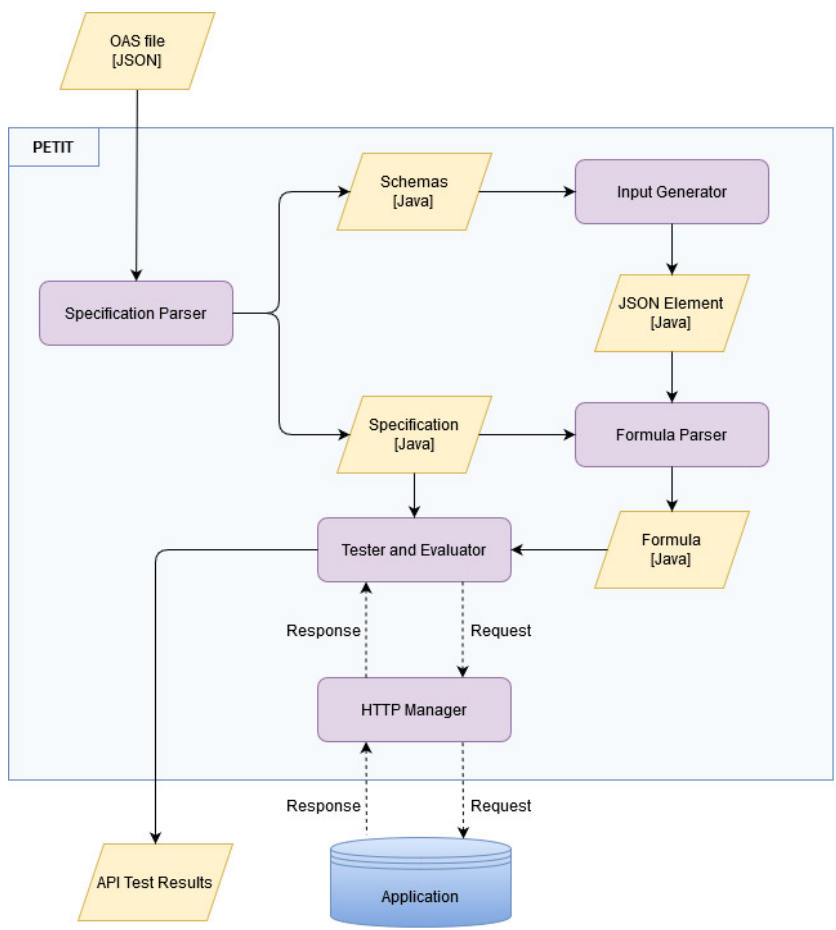

Figure 4.6: PETIT's architecture.

As previously mentioned, PETIT can be executed with the following four parameters, only two of them being mandatory:

**File Path** the complete path to the JSON file containing the OAS document.

**Operation Order Strategy** API's operations are categorized into Constructors, Mutators and Observers. The order strategy is the order in which these operations' categories will be tested. The operation order within each category is random. Hereupon, a valid strategy would be, e.g., CMO where the constructors would be tested first, then the mutators and, finally, the observers. Operations can also be tested randomly by providing RND as the strategy. When this parameter is wrongly specified the message in listing 4.5 is displayed.

```
Invalid operation order strategy.
A valid strategy is composed of three characters meaning the following:
   > C: constructors (POST)
   > M: Mutators (PUT, DELETE)
   > O: Observers (GET)
   > RND (random)
A valid strategy would be, e.g., CMO
```

Listing 4.5: Error message when operation order strategy is wrongly specified.

**Verbose Mode (-v)** if this flag is present, all performed requests' responses will be shown. This mode is accompanied by another argument which indicates the number of resources to be printed.

**Random API Order (-r)** if this flag is present, the APIs described in the specification will be shuffled and tested in a random order.

Both the file path and operation order strategy parameters are required. The remaining are not required and, therefore, the order in which they are specified is irrelevant.

PETIT's output is a detailed description of the testing process results. It comprises detailed information on what is happening during each stage of the testing process, while testing each operation. When an API test is complete the number of succeeded, failed, and inconclusive tests are shown. Since PETIT is making changes to the microservice's database it also reverts all changes when the test process is finished. This cleanup is particularly important since PETIT only generates valid input data and, if not removed, besides wasting memory, it may cause, e.g., a tournament to be full when, in fact, it is full with dummy players. Listing 4.6 shows PETIT's output when testing an API with a single operation.

```
>>> Testing POST /players
  > Verifying Invariants                    :  OK
  > Generating Data                         :  OK
  > Verifying Preconditions                 :  OK
  > Performing Request                      :  OK
  > Verifying Postconditions                :  OK
  -------------------------------------------------------
  POST /players                             :  OK

  -------------------------------------------------------
>>> Player's API Results:
OK                                          :  1
NOT OK                                      :  0
INCONCLUSIVE                                :  0

>>> REVERTING ALL EFFECTS                   :  OK
```

Listing 4.6: PETIT's output when testing an API with a single operation.

With all this information in mind, one possible way of executing PETIT is depicted in listing 4.7. This would execute PETIT in verbose mode (showing maximum of two resources), with random API order and MCO (mutators, constructors and observers) strategy.

```
$ java -jar PETIT.jar openapi.json CMO -v -r

>>> Maximum resources to be printed: 2
```

Listing 4.7: PETIT's output when testing an API with a single operation.

This chapter provided the core concepts to understand both APOSTL's and PETIT's design process. The next chapters will present an implementation as well as its limitations.

## SOLUTION IMPLEMENTATION

This chapter presents essential information on how PETIT and APOSTL are implemented. The specification language implementation section illustrates how the Open API Specification extension and how APOSTL's integration with PETIT were achieved, as well as a formal definition for APOSTL's grammar and its restrictions.

The testing tool implementation section describes the most relevant aspects of PETIT's implementation, namely a detailed description of all its architectural components, the testing process it implements, and the detailed process for valid test data generation.

## 5.1  Specification Language: APOSTL

As previously mentioned, APOSTL is a specification to annotate APIs' specifications with useful contracts for testing purposes, based on first-order logic with some restrictions. This section aims to expose the needed steps to implement APOSTL, namely how the extension of Open API Specification is achieved, a formal description of APOSTL's rules, and APOSTL's restrictions.

### 5.1.1  Extending OpenAPI Specification

Open API Specification allows the addition of custom properties to a specification description. In order to accommodate APOSTL's conditions in an OAS document, there were added three new properties: *x-requires* for the preconditions, *x-ensures* for the post-conditions, and *x-invariants* for the invariants. It was also added a fourth property to aid in custom test data generation, *x-regex*. This last property can be found in schemas descriptions such as in operations' parameters schemas and model schemas.

The properties representing operations' contracts – *x-requires* and *x-ensures* –, and the property representing API invariants – *x-invariants* – are collections, meaning they can have more than one APOSTL condition. On the other hand, *x-regex* property can only comprise a single regular expression.

As seen in section 2.5.3, the OAS document has a well defined structure. Although custom properties can be added anywhere in the document, their position could interfere in readability and usability. As such, the main concern was where should the new properties be added so that its position is not disturbing and is easy to understand to which operation, or API, do they belong to. Returning to the tournaments' application description, listing 5.1 depicts the partial description of the operation responsible for player deletion. As seen in the listing, *x-requires* and *x-ensures*, concerning operations, appear in the beginning of an operation description, right after its summary. When the operation has a parameter, the information concerning the parameter generation, *x-regex*, appears within the parameter schema description, also depicted in listing 5.1.

```
1  "/players/{playerNIF}":
2    delete:
3      summary: Delete the player with the given NIF.
4      x-requires:
5      - response_code(GET /players/{playerNIF}) == 200
6      x-ensures:
7      - response_code(GET /players/{playerNIF}) == 404
8      - response_body(this) ==
9        previous(response_body(GET /players/{playerNIF}))
10     parameters:
11     - name: playerNIF
12       schema:
13         type: string
14         x-regex: "(1|2)[0-9]{8}"
```

Listing 5.1: YAML object for Player's API delete player operation.

Invariants are conditions concerning APIs and, as such, they appear in the beginning of APIs' descriptions. Listing 5.2 shows the beginning of the tournament's API description and where the its *x-invariants* property is located.

```
1  "/tournaments":
2    x-invariants:
3    - for t in response_body(GET /tournaments) :-
4      response_body(GET /tournaments/{t.tournamentId}/enrollments).length
5      <= response_body(GET /tournaments/{t.tournamentId}/capacity)
```

Listing 5.2: YAML object for Tournament's API.

With this implementation every new property is as close as possible to what relates to without, at the same time, being too intrusive hampering usability.

| | | |
|---|---|---|
| `formula` | ::= | `quantifiedFormula` \| `booleanExpression` |
| `quantifiedFormula` | ::= | `quantifier` string in `call :- booleanExpression` |
| `quantifier` | ::= | for \| exists |
| `call` | ::= | `operation` \| `operationPrevious` |
| `booleanExpression` | ::= | `booleanExpression booleanOperator booleanExpression` \| `clause` |
| `clause` | ::= | T \| F \| `comparison` |
| `comparison` | ::= | `term comparator term` |
| `term` | ::= | `operation` \| `operationPrevious` \| `param` |
| `operationPrevious` | ::= | previous ( `operation` ) |
| `operation` | ::= | `operationHeader` ( `operationParameter` ) function? |
| `operationHeader` | ::= | request_body \| response_body \| response_code |
| `operationParameter` | ::= | `httpRequest` \| this |
| `httpRequest` | ::= | `method` \| `url` |
| `url` | ::= | `segment`+ |
| `method` | ::= | GET \| POST \| PUT \| DELETE |
| `comparator` | ::= | == \| != \| <= \| >= \| < \| > |
| `booleanOperator` | ::= | && \| \|\| \| => |
| `param` | ::= | string (. string)* \| int |
| `segment` | ::= | / `block`(. block)* |
| `block` | ::= | { `blockParameter` } \| string |
| `blockParameter` | ::= | string (. string)? \| `operation` \| `operationPrevious` |
| `function` | ::= | . string |

Table 5.1: APOSTL's grammar defined in BNF.

## 5.1.2 Grammar

APOSTL's grammar is a context-free grammar, meaning its non-terminal rules can be applied regardless of the context it is inserted, meaning the left hand side of a non-terminal rule can always be replaced by the right side of the same rule, independently of the circumstances where this rule appears.

Backus-Naur form (BNF) is a commonly used notation for describing grammars. Every rule in BNF has the following structure:

*rule_name ::= expansion*

An expansion may contain terminal and non-terminal rules. These rules are connected either by alternatives or sequences. APOSTL's grammar is described in table 5.1. Terminal symbols are depicted in blue for readability purposes.

An APOSTL formula can either be a boolean expression or a quantified formula. An example of an APOSTL quantified formula can be found in tournament's API invariant, as seen in listing 5.2. A boolean expression is recursively defined as being two boolean expressions, separated by a boolean operator, or a clause. In turn, a clause can either be a boolean value – true (T) or false (F) –, or a comparison, which is made up of two terms, that can either be APOSTL operations or parameters, and a comparator. An example of an APOSTL comparison can be found in listing 5.1, which shows a player's API operation contract.

### 5.1.3 Integration with PETIT

In order for PETIT to be able to evaluate APOSTL's formulas, there is the need to tell whether a formula is formed according to APOSTL's rules, i.e., its grammar. Hereupon, there is the need to implement a parser, a program that analyses a sequence of tokens and checks if this sequence is conforming to the grammar.

Instead of implementing a parser from scratch, PETIT uses a tool to generate it. ANTLR – ANother Tool for Language Recognition – is a parser generator that, given a formal language description, can automatically build and traverse parse trees [29]. Parse trees are data structures that can be traversed in order to tell whether the input matches the grammar. A parse tree resulting from running the parser generated by ANTLR with the formula *response_code(GET /players/{playerNIF}) == 404* is depicted in figure 5.1.

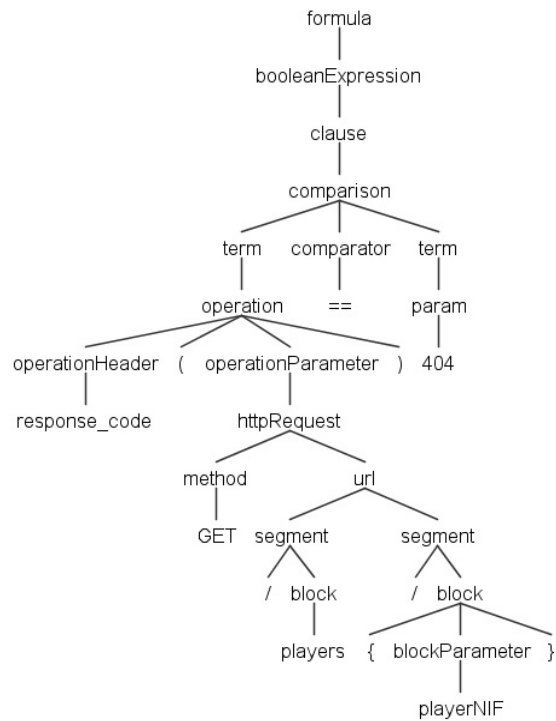

Figure 5.1: Parse tree of a conforming APOSTL formula.

When a formula is not conforming to the grammar rules, ANTLR throws an exception which is, in turn, caught and handled by PETIT.

Integration of APOSTL with PETIT involves not only traversing the parsing tree and checking formulas' conformity to the grammar, but also evaluating APOSTL's formulas with the generated input. This will be further analysed in the following section, namely when describing PETIT's component *formula parser*.

### 5.1.4 Restrictions

By analysing APOSTL's grammar, described in table 5.1, and as previously referred, APOSTL does not support nested quantifiers, as depicted in listing 5.3, neither quantifiers with more than one variable, as depicted in listing 5.4.

```
for t in response_body(GET /tournaments) :-
  for p in response_body(GET /tournaments/{t.tournamentId}/players) :-
    response_code(/tournaments/{tournamentId}/enrollments/{p.playerNIF} == 200
```

Listing 5.3: A nested quantifier, written in APOSTL.

```
for t in response_body(GET /tournaments),
  p in response_body(GET /tournaments/{t.tournamentId}/players) :-
    response_code(/tournaments/{tournamentId}/enrollments/{p.playerNIF} == 200
```

Listing 5.4: A quantifier with more than one variable, written in APOSTL.

Both these conditions mean the exact same: for every tournament if a player is stored in the tournament's *players* collection, the player must be enrolled in the tournament.

There are some restrictions in APOSTL's implementation which, by only analysing its grammar, could be considered allowed. According to the grammar's rules an HTTP operation can be a GET, POST, PUT or DELETE. However, and as previously referred, APOSTL's formulas can only be made up of pure HTTP operations, meaning only GET operations can be used. It is also not allowed for the keyword *this* to appear anywhere else but in *comparisons*. In other words, *this* cannot appear in a *quantified formula*'s *call*. Also contrary to what is described in the grammar, composed *block parameters* can only have depth one, meaning that *block parameters* such the one depicted in listing 5.5 cannot occur, since it has depth two (*p.playerNIF.tournaments*).

```
for p in request_body(GET /players) :-
  response_code(GET /players/{p.playerNIF.tournaments}) == 200
```

Listing 5.5: An invalid block parameter in an APOSTL's formula, according to its implementation.

Although APOSTL's grammar does not have any information about *x-regex* parameters, its implementation assumes that schemas cannot have a composed identifier, meaning each resource can only have one property as its ID. This happens for no particular reason other than lack of time.

APOSTL's implementation also assumes that properties that serve as IDs cannot have the same name in different resources. In short, different properties belonging to different resources must have different names. This happens to prevent having to specify the resource type in order to get its ID, i.e., if both players and tournaments resources would have its identification property named *id*, there would be the need to refer to them as *t.id* and *p.id* – instead of just *tournamentId* and *playerNIF* – and, consequently, having to define *p* as a player and *t* as a tournament in APOSTL specifications.

## 5.2 Testing Tool: PETIT

PETIT is a tool which automates the microservice testing process based on its API description. This section aims to illustrate PETIT's implementation from its architectural components to the implemented testing process.

### 5.2.1 Architecture Components

PETIT's overall architecture is shown in figure 4.6. It illustrates all PETIT's components – *specification parser*, *input generator*, *formula parser*, *tester and evaluator*, and the *HTTP manager* – as well as their interactions. All these components are responsible for performing a different, but equally, important task. As such, their implementation and interactions will be further analysed.

**Specification Parser** as the name implies, this component is a parser responsible for analysing and translating the OAS document. From a JSON specification, it generates a Java object with all the information in the OAS file, and several Java objects, one for each schema.

**Input Generator** is responsible for all test data generation. The *generator* operation, depicted in figure 5.2, begins by checking the operation type – POST, PUT, GET or DELETE. If the operation is a POST or a PUT, it generates a JSON object form the operation's body schema, depicted in figure 5.3. Otherwise, i.e., if it is a GET or a DELETE and the operation has parameters, the JSON object is generated form the URL parameter description, depicted in figure 5.4.

*Generate form body schema* operation, illustrated in figure 5.3, starts by going through all operation's properties. For each property type there is a different outcome. If the property is a string and, simultaneously, a database generated property then there is no need to generate it. A flag indicated the property is generated is added to the object being generated. If the property is a string that is not database generated, then if it has a regular expression, the string will be generated according to the regular expression; otherwise a random string is generated. If the property is an integer and is database generated, the process is the same as described for string properties. If it is not database generated and it has a minimum value, the integer will be generated according to that minimum value, ranging from the minimum up until the maximum integer. If the minimum value is not present, then a random positive integer is generated. For properties of the type array an empty one is generated. For object properties, the *generate from body schema* operation is called recursively.

*Generate from URL parameter* operation, illustrated in figure 5.4, begins by checking if the parameter type is string or integer. In the case of being a string, then the parameter is generated from the regular expression. Otherwise, the integer is generated ranging from the specified minimum to the maximum integer.

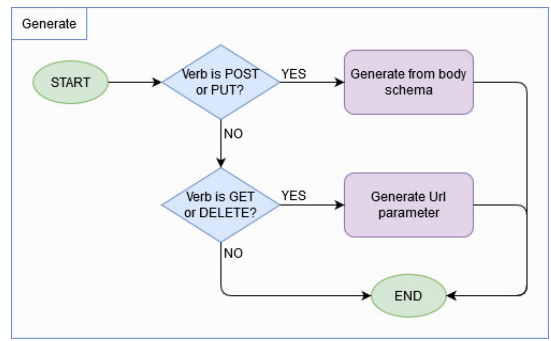

Figure 5.2: Generate operation logic.

**Formula Parser** component is responsible for traversing the parsing tree that is generated by ANTLR. Each node of the parsing tree needs to be checked in order to ascertain if a formula is conforming to the grammar's rules. The *Visitor Oriented Parser* was developed for that purpose, based on [41]. The visitor design pattern has the purpose of separating an algorithm from the object it operates on. It allows to add new functionality to an already implemented class without changing its implementation. A visitor usually operates in a class that is composed by several other element classes. In APOSTL's case, the *formula* class is composed by several element classes such as *boolean expression*, *quantified formula*, and so forth.

**HTTP Manager** as the name implies, it is responsible for the HTTP request and response management. HTTP responses are parsed into Java objects so they can be easily manipulated.

**Tester and Evaluator** has the purpose of implementing the testing process, described in subsection 5.2.2, managing the generated objects' pool, and evaluating all APOSTL formulas. The object pool is a mechanism implemented in order to enhance PETIT's performance. Every time new test data is generated it is added to the pool. When data of the same type is needed for another test, instead of generating new data, the pool is checked and, if there is conforming data, it gets recycled.

An evaluation consists of ascertain the truth value of an APOSTL formula. Algorithm 1 depicts how a quantified formula is evaluated. It starts by retrieving the quantified formula's collection from the database. For each element in the collection, the boolean expression's URL parameters are replaced for the element's values. Then, the resulting boolean expression is evaluated, and its result is stored. If the formula has the universal quantifier, for the first element that this evaluation result is false, the quantified formula also evaluates to false. Otherwise, if the formula is quantified by the existential quantifier, for the first element that the partial evaluation is true, the quantified formula also evaluates to true.

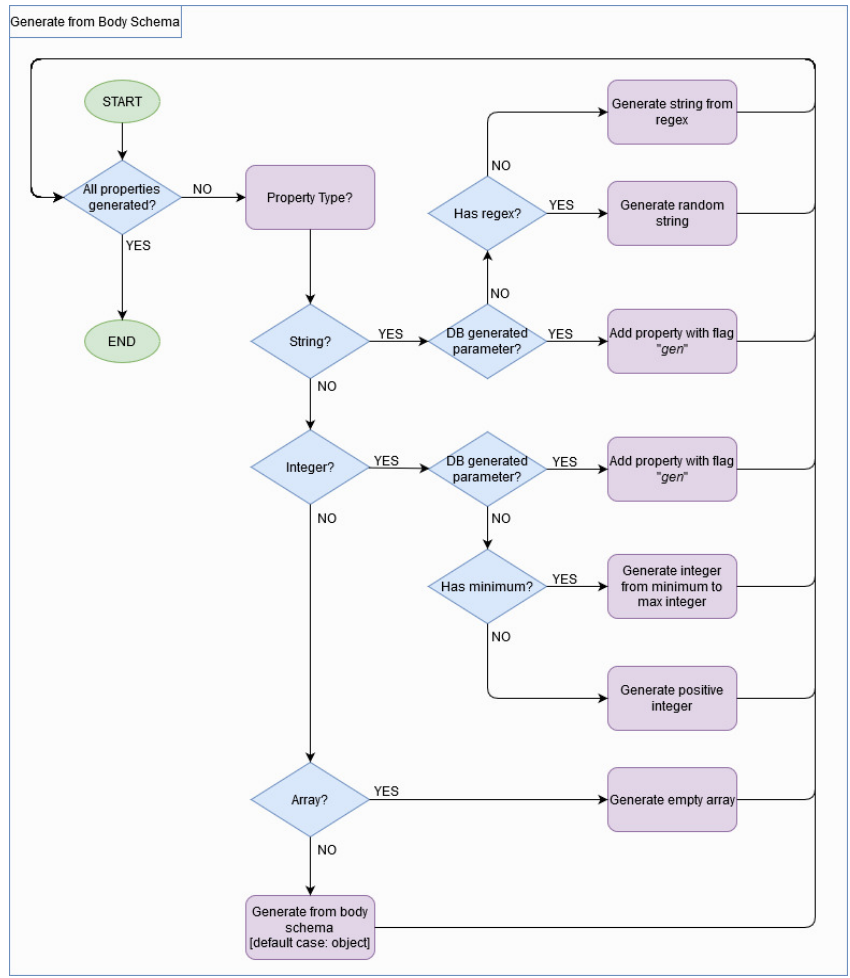

Figure 5.3: Generate body schema operation logic.

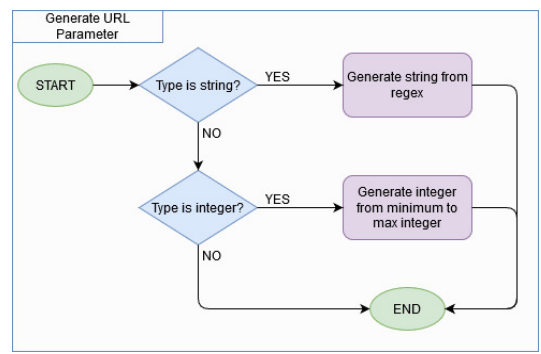

Figure 5.4: Generate URL parameter operation logic.

---

**Algorithm 1** Evaluation of ALPOSTL quantified formulas.

    ▷ Evaluates a quantified formula.
1: **function** EVALUATEQUANTIFIED(parser, formula)
2:    isUniversal ← formula.isUniversal()
3:    booleanExpression ← formula.getExpression()
4:    collectionURL ← formula.getCollectionUrl()
5:    collection ← HTTPManager.GET(collectionURL)    ▷ perform GET request
6:    **for** elem ∈ collection **do**
7:        parameters ← getConditionURLParameters(booleanExpression)
8:        **for** p ∈ parameters **do**
9:            booleanExpression ← replaceURLParameters(booleanExpression, p, elem)
10:      f ← parser.parse(formula)    ▷ transform string into formula obj
11:      partialResult ← evaluateFormula(f)    ▷ evaluate the current expression
12:      **if** isUniversal **then**    ▷ for the first elem that eval is false return false
13:          **if** !partialResult.getValue() **then**
14:             **return** false
15:      **else**    ▷ for the first elem that eval is true return true
16:          **if** partialResult.getValue() **then**
17:             **return** true

---

### 5.2.2 Testing Process

The testing process implemented by PETIT has three core operations, decreasing in granularity: *testSpec*, *testAPI* and *testOperation*.

The *testSpec* implementation is depicted in algorithm 2. It starts by checking if the user provided the *r* flag which, if it is present, means the APIs' testing order will be randomized. After this check, the operation enters a loop testing all APIs, either in the randomized order or the original order in which they are defined in the OAS file. When all APIs are tested, all the changes made to the microservice database are reverted by gathering all operations responsible for resource deletion and performing them on every object in the object pool, which concludes the specification testing process.

The *testAPI* implementation is depicted in algorithm 2. The process starts by reorganizing all API's operations into the order that was specified by the user – e.g. CMO (constructors, then mutators and, finally, observers). Similarly to the previous operation, it enters a loop verifying the API's invariants and testing all operations, by the previously defined order. When all operations are tested, the API testing results are shown and the API testing process is complete.

Finally, *testOperation*, depicted in algorithm 2, is responsible for testing each individual operation. This testing step can be divided into two sections: the test data generation logic and the operation testing per se.

**Algorithm 2** Algorithm for testing a specification and its main functions.

      ▷ Tests a specification.
1: **function** TESTSPECIFICATION(spec)
2:     APIs ← spec.getAPIs()
3:     apiResults ← ∅
4:     **for** api ∈ APIs **do**
5:         apiResults ← testAPI(api)
6:         printAPIResults(apiResults)
7:     deleteEffects(spec.getDeletes())

      ▷ Tests a single API.
8: **function** TESTAPI(api, strategy)
9:     operations ← reorganize(api.getOperations(), strategy)
10:    apiResults ← ∅
11:    **for** op ∈ operations **do**
12:        satisfiesInvariants(api)
13:        apiResults.add(testOperation(op))
14:    **return** apiResults

      ▷ Tests an API operation.
15: **function** TESTOPERATION(op)
16:    verb ← op.getVerb()
17:    url ← op.getUrl()
18:    params = getURLParameters(url)
19:    **if** verb ≠ POST **then**
20:        generated ← recycle(params)
21:        **if** generated = null **then**
22:           generated ← generate(op)
23:    **else**
24:        generated ← generate(op)
25:        addToPools(op)
26:    url ← replaceParameters(params)
27:    satisfiesPre ← processPreconditions(op, generated, generatedURLParam)
28:    previousResults ← processPrevious(op, generatedURLParam, generated)
29:    response ← performRequest(op, url, generated)         ▷ operation's request
30:    **if** verbose **then**           ▷ executed in verbose mode
31:        printResponse(response)
32:    **if** res.getCode() ≠ 200 **then**
33:        printCausedBy(response)
34:    **else**
35:        satisfiesPos ← processPostconditions(op, generated, response)
36:        satisfiesPrev ← satisfiesPrevious(op, generated, response)
37:    opOk ← response.getCode() = 200 ∧ satisfiesPre ∧ satisfiesPos ∧ satisfiesPrev
38:    failedAsExpected ← res.getCode() ≠ 200 ∧ ¬satisfiesPre
39:    analyse ← res.getCode() ≠ 200 ∧ satisfiesPre
40:    result ← getOperationResult(opOk, failedAsExpected, analyse)
41:    printOperationResult(op, opOk, failedAsExpected, analyse)
42:    **return** result

The test data portion starts by checking if the operation is a constructor, i.e. a POST. If it is, new test data is generated. Otherwise, the generated objects' pool is checked. If it is empty, then new test data is generated. If it has some previously generated elements and there is at least one element which has the same schema as the element needed to perform the operation, then this element is recycled, meaning it will be used again for this operation's test. If there is no element with the same schema, a new element is generated.

When the testing data is set, either by recycling or generation, there is the need to replace the URL parameters – including the operation URL and all pre and postconditions – with the correct values taken from the element's properties. The replacement operation implementation is described in algorithm 3. When every parameter is replaced by the correct values the testing process begins. It starts by verifying if the generated element is conforming to the preconditions, depicted in algorithm 3. If not, the failed preconditions are displayed and the testing process is resumed, in order to check the microservice's response. Otherwise, it will search for postconditions with the *previous* keyword and, if there are some, they are processed, meaning all its requests are performed; if not, the testing process continues by performing the operation's request. In case the user executed PETIT in verbose mode – *v* flag is present –, then the request's response will be displayed. If the request failed, all the known reasons why it failed are displayed, the operation testing results are also displayed and the testing process ends. Otherwise, i,e, if the request does not fail, the operation's postconditions are verified – depicted in algorithm 3 – taking the response and the generated data into account. If a postcondition fails it is displayed. Postconditions with the *previous* keyword are now verified – taking into account their results were obtained before the operation request was performed. If there are some failed postconditions with the *previous* keyword, they also get displayed. The operation testing results are displayed and the operation testing process is complete.

This chapter described both PETIT's and APOSTL's implementation. The next chapter aims to point some additional aspects by using PETIT with two different applications: a correct, and a faulty one.

---

**Algorithm 3** Auxiliary operations: evaluating contracts and replacing parameters.

---

      ▷ Evaluates preconditions and processes its output.
1: **function** PROCESSPRECONDITIONS(op, generated, generatedURLParam)
2:     failedPreconditions ← satisfiesPRE(op, generated, generatedUrlParam)
3:     satisfiesPre ← failedPreconditions = ∅ ? true : false
4:     **if** !satisfiesPrev **then**
5:         printFailedConditions(failedPreconditions)
6:     **return** satisfiesPre

      ▷ Evaluates postconditions and processes its output.
7: **function** PROCESSPOSTCONDITIONS(op, generated, response)
8:     ensures ← removePrevious(op.getEnsures())
9:     failedPostconditions ← satisfiesPOS(ensures, generated, response)
10:     satisfiesPos ← failedPostconditions = ∅ ? true : false
11:     **if** !satisfiesPos **then**
12:         printFailedConditions(failedPostconditions)
13:     **return** satisfiesPos

      ▷ Evaluates postconditions with the *previous* keyword and processes its output.
14: **function** SATISFIESPREVIOUS(op, generated, response)
15:     **if** previousResults ≠ ∅ **then**
16:         failedPrevious ← evaluatePrevious(previousResults, response)
17:     satisfiesPrev ← failedPrevious = ∅ ? true : false
18:     **if** !satisfiesPrev **then**
19:         printFailedConditions(failedPrevious)
20:     **return** satisfiesPrev

      ▷ Replaces URL parameters for generated values.
21: **function** REPLACEPARAMETERS(parameters, url)
22:     **if** parameters ≠ ∅ **then**
23:         **for** param ∈ parameters **do**
24:             poolElem ← findObject(param)         ▷ checks if the pool has usable obj.
25:             **if** poolElem ≠ null **then**
26:                 url ← replaceURLParameters(url, param, poolElem.get(param))
27:             **else**                 ▷ generate parameter from regex or min
28:                 regex ← spec.getParameterRegex(param)
29:                 min ← spec.getParameterMin(param)
30:                 type ← spec.getParamType(param)
31:                 generatedURLParam ← generateURLParam(type, min, regex)
32:                 url ← replaceURLParameters(url, param, generatedURLParam)
33:     **return** url

---

# 6

## EVALUATION

As previously discussed, PETIT can be executed with different operation order strategies. Different strategies can lead to different test outcomes. Hereupon, this chapter features several tests conducted on tournaments' application, described in section 4.1, to ascertain how the order strategy parameter influences the test result. Each of the following sections illustrate how the different operation categories – constructors, observers and mutators – can be tested both for success and failure cases. Recalling the application's description, one knows that it is made up of two different APIs – the players and the tournaments API. PETIT sequentially tests each APIs' operations in the specified order. PETIT is not executed in random mode – *r* flag –, so players' API is always tested first. For readability purposes, this chapter's listings only depict non-trivial or error cases, and the order in which each operation appears is the order in which it is tested.

This chapter analyses PETIT's tests results when testing a correct implementation of the tournaments' application as well as a faulty one. Implementation errors will be incrementally added in order to ascertain if PETIT finds them and, if it does, how useful is its output.

## 6.1 Testing Constructors

The most adequate order strategies to test constructor operations for their success case – the used test data is conforming to the constructors' contract – are COM and CMO. Both this strategies test constructors first, meaning the following operations being tested use the resources created by the constructors. If constructors have some implementation error, it will likely be caught in the following tests. Assuming constructors are implemented according to its specification, both this strategies can also be used to test mutators and observers for the success case. On the other hand, if one assumes constructors are not implemented according to its specification, both observers and mutators will be tested for their failure scenarios.

Listing 6.1 shows the specification testing results when testing it with COM order strategy. Although everything appears to be correct, there is always the need to check the execution trace, i.e, each operation's testing output.

```
>>> Player's API Results:
    OK                                    :  6
    NOT OK                                :  0
    INCONCLUSIVE                          :  0
------------------------------------------------------------------------
>>> Tournament's API Results:
    OK                                    :  10
    NOT OK                                :  0
    INCONCLUSIVE                          :  0
```

Listing 6.1: Specification test results when executing PETIT with COM order strategy.

Listing 6.2 shows PETIT's output, when performing the same test, at operation level. One can see that, besides producing a result that is still considered correct, there were three operations that were not tested for the success case: inserting, retrieving and removing an enrollment. In listing 6.2 the result of inserting a new enrollment is classified as *failed (as expected)*. This happens because some preconditions did not hold before the request was made. Considering the first operation in the same listing – inserting a new enrollment – one can see that the operation failed because neither the player nor the tournament exist in the system and, therefore, a new enrollment could not be added. Since player's API was tested first, there should be, at least, one player stored in the pool. Recalling the testing process, described in section 5.2.2, one knows that every correctly generated object is stored in the data pool. The player is, in fact, stored in the data pool and recycled to test the enrollment insertion operation. However, the player's API was tested first, meaning the player deletion operation was previously tested as well. Therefore, although being stored in the data pool, if the player deletion operation is correctly implemented the player will not be stored in the microservice's database.

The result of the operation responsible for retrieving an enrollment is also labeled as *failed (as expected)*. This time, the only failing precondition is the one concerning the player, for the reason previously described. Since the strategy chosen is COM, there is already a tournament in the system that was not yet deleted – constructors are tested before mutators.

The last operation failing, as expected, is the enrollment deletion. This is the last API operation being tested and, as such, the failing preconditions concern both the player and the tournament that were already deleted, and the enrollment that ended up not being created in the first place.

This test case shows that, even though PETIT labels the specification test as being successful, not all possible operations' outcomes are, in fact, being tested. Hereupon, there is the need to test the same application with different strategies in order to increase test coverage. However, since the system under test is a black box, test coverage cannot be effectively measured – in the sense of lines of code or conditional branches covered. In a black box testing scenario the applications' end-user play a large role of determining the test coverage and, therefore, cannot be measured accurately.

```
>> POST /tournaments/{tournamentId}/enrollments
  > Verifying Invariants                     :  OK
  > Generating Data                          :  OK
  > Verifying Preconditions                  :  NOT OK
    > Failed:
      - response_code(GET /tournaments/31) == 200
      - response_code(GET /players/223893138) == 200
  > Performing Request                       :  FAILED (as expected)
    > Caused by:
      > Code: 404
      > Message: Player with NIF 223893138 not found.
  ------------------------------------------------------------------------
  POST /tournaments/{tournamentId}/enrollments  :  OK

  >> GET /tournaments/{tournamentId}/enrollments/{playerNIF}
  > Verifying Invariants                     :  OK
  > Recycling Data                           :  OK
  > Verifying Preconditions                  :  NOT OK
    > Failed:
      - response_code(GET /players/223893138) == 200
  > Performing Request                       :  FAILED (as expected)
    > Caused by:
      > Code: 404
      > Message: Player with NIF 223893138 does not exist.
  ------------------------------------------------------------------------
  GET /tournaments/{tournamentId}/enrollments/{playerNIF} :  OK

>> DELETE /tournaments/{tournamentId}/enrollments/{playerNIF}
  > Verifying Invariants                     :  OK
  > Recycling Data                           :  OK
  > Verifying Preconditions                  :  NOT OK
    > Failed:
      - response_code(GET /tournaments/2) == 200
      - response_code(GET /players/223893138) == 200
      - response_code(GET /tournaments/2/enrollments/223893138) == 200
  > Performing Request                       :  FAILED (as expected)
    > Caused by:
      > Code: 404
      > Message: Player with NIF 223893138 does not exist.
  ------------------------------------------------------------------------
  DELETE /tournaments/{tournamentId}/enrollments/{playerNIF} :  OK
```

Listing 6.2: PETIT's partial output of a tournaments' API test executed with COM strategy.

With the COM order strategy, one can effectively test constructor and observer methods. However, since tournaments' API has more than one constructor, the order in which each constructor is tested will also have an effect on the test outcome. If the constructor enrolling a new player in a tournament is tested first, there will be no tournament in the system, therefore, it will fail. If the order is reversed, i.e. the tournament constructor is tested first, the test success will only depend on the player being stored in the microservice data base. These limitations will be further addressed in the next chapter, namely when discussing the improvement possibilities and the future work.

Listing 6.3 depicts the tournaments' application testing results when testing it with CMO order strategy. Just like in the previous test, there are several operations whose test result is *failed (as expected)*, namely, the operation responsible for updating a tournament resource. This happens as a result of the tournament deletion being tested before the tournament update and, consequently, the tournament does not exist in the system.

```
>>> Player's API Results:
    OK                                          :  6
    NOT OK                                      :  0
    INCONCLUSIVE                                :  0
------------------------------------------------------------------------
>>> Tournament's API Results:
    OK                                          :  9
    NOT OK                                      :  0
    INCONCLUSIVE                                :  1
```

Listing 6.3: Specification test results when executing PETIT with CMO order strategy.

By analysing PETIT's output, one can see that there is one operation whose test is inconclusive. Through analysing each operations' output, the inconclusive operation test is identified, and depicted in listing 6.4. In this case, the operation responsible for retrieving a tournament fails even though all preconditions hold. This happens as a result of mutators being tested before observers, and the tournament deletion operation being implemented according to its specification. Therefore, trying to retrieve the tournament that was previously deleted will result in the tournament not being found, which, in this case, is considered the correct behaviour.

```
>> PUT /tournaments/{tournamentId}
  > Verifying Invariants                        :  OK
  > Recycling Data                              :  OK
  > Verifying Preconditions                     :  NOT OK
    > Failed:
      - response_code(GET /tournaments/2) == 200
  > Performing Request                          :  FAILED (as expected)
    > Caused by:
      > Code: 404
      > Message: Tournament with id 2 not found.
```

```
-------------------------------------------------------------------------
  PUT / tournaments /{ tournamentId }                   :  OK

>> GET / tournaments /{ tournamentId }
 > Verifying Invariants                                 :  OK
 > Recycling Data                                       :  OK
 > Verifying Preconditions                              :  OK
 > Performing Request                                   :  FAILED ( analyse exec. trace )
   > Caused by :
     > Code : 404
     > Message : Tournament with id 2 not found .
-------------------------------------------------------------------------
  GET / tournaments /{ tournamentId }                   :  INCONCLUSIVE
```

Listing 6.4: PETIT's partial output of a tournaments' API test executed with CMO strategy.

As previously referred, both this strategies can be used to test mutator and observer operations. As such, CMO strategy can be used to test mutators and COM can also be used to test observers.

In the first testing scenario, although the specification test results are positive, by looking into each operation test result, one can conclude that not all possible outcomes were tested. In the second testing scenario, on the other hand, there is an inconclusive test case that is not, necessarily, wrong. Ultimately, what both these scenarios aim to enforce is that one should perceive PETIT's output in a critical perspective, not only looking into the specification test results as a whole, but also into each operation result and the order in which they were tested.

## 6.2 Testing Mutators

Testing mutators for its success case will fall into the previously discussed order strategy, CMO. This happens because in order for mutator operations to perform correctly they need to work on previously existing resources. This means that, assuming constructors and observers are correctly implemented, mutators input will be correctly defined and its effects will be noticeable when testing observers. However, there is still the need to test these operations when the test data is not conforming to their contract. PETIT is able to do this when provided with MCO or MOC order strategies. Testing the tournaments' application specification with MCO order strategy produces the same results as the ones shown in listing 6.3.

Listing 6.5 depicts player's API mutator operations' results. Since mutator operations are the first to be tested, there is no data to be updated nor removed. As seen on listing 6.5, the preconditions for both operations – updating and removing a player – fail. Since tournaments' application is implemented according to its specification, the request fails, as expected, and the operations' testing results are positive.

```
>> PUT /players/{playerNIF}
  > Verifying Invariants                      :   OK
  > Recycling Data                            :   OK
  > Verifying Preconditions                   :   NOT OK
    > Failed:
      - response_code(GET /players/212145124) == 200
  > Performing Request                        :   FAILED (as expected)
    > Caused by:
      > Code: 404
      > Message: Player with NIF 212145124 not found.
---------------------------------------------------------------------
  PUT /players/{playerNIF}                    :   OK

>> DELETE /players/{playerNIF}
  > Verifying Invariants                      :   OK
  > Recycling Data                            :   OK
  > Verifying Preconditions                   :   NOT OK
    > Failed:
      - response_code(GET /players/270771533) == 200
  > Performing Request                        :   FAILED (as expected)
    > Caused by:
      > Code: 404
      > Message: Player with NIF 270771533 not found.
---------------------------------------------------------------------
  DELETE /players/{playerNIF}                 :   OK
```

Listing 6.5: PETIT's partial output of a players' API test executed with MCO strategy.

The tournaments' API mutators operations' testing results are similar to the ones of players' API. However, listing 6.3 shows that there was an inconclusive test for a tournaments' API operation. The operation whose test is inconclusive is the one responsible for checking whether a player is enrolled in a tournament. By analysing the test sequence, shown in listing 6.6, the reason is clear: the operation responsible for inserting an enrollment was tested first, meaning there was still no tournament stored in the system; the execution proceeds with inserting a tournament and then with checking if a player is enrolled in the tournament that was just inserted. PETIT classifies this test as inconclusive because it lacks information about the execution trace. By analysing it, one can state that the microservice behaviour was, in fact, correct.

By being able to detect the previously described test case, one can conclude that this order strategy could simultaneously be used to test constructor operations.

Listing 6.7 shows the results of testing the tournaments' application with MOC order strategy. As seen in the listing, both player's and tournament's APIs have one inconclusive operation test.

```
>> POST /tournaments/{tournamentId}/enrollments
 > Verifying Invariants                 :  OK
 > Generating Data                      :  OK
 > Verifying Preconditions              :  NOT OK
   > Failed:
     - response_code(GET /tournaments/46) == 200
 > Performing Request                   :  FAILED (as expected)
   > Caused by:
     > Code: 404
     > Message: Tournament with ID 46 not found.
-----------------------------------------------------------------------
 POST /tournaments/{tournamentId}/enrollments :  OK

>> POST /tournaments
 > Verifying Invariants                 :  OK
 > Generating Data                      :  OK
 > Verifying Preconditions              :  OK
 > Performing Request                   :  OK
 > Verifying Postconditions             :  OK
-----------------------------------------------------------------------
 POST /tournaments                       :  OK

>> GET /tournaments/{tournamentId}/enrollments/{playerNIF}
 > Verifying Invariants                 :  OK
 > Recycling Data                       :  OK
 > Verifying Preconditions              :  OK
 > Performing Request                   :  FAILED (analyse exec. trace)
   > Caused by:
     > Code: 404
     > Message: Player with NIF 220810071 is not enrolled in the tournament 2.
-----------------------------------------------------------------------
 GET /tournaments/{tournamentId}/enrollments/{playerNIF}  :  INCONCLUSIVE
```

Listing 6.6: PETIT's partial output of a tournaments' API test executed with MCO strategy.

```
>>> Player's API Results:
    OK                                  :  5
    NOT OK                              :  0
    INCONCLUSIVE                        :  1
-----------------------------------------------------------------------
>>> Tournament's API Results:
    OK                                  :  9
    NOT OK                              :  0
    INCONCLUSIVE                        :  1
```

Listing 6.7: Specification test results when executing PETIT with MOC order strategy.

The operations whose test result is inconclusive are the ones responsible for retrieving a player and a tournament resource. Since the PETIT is executed with MOC, the observer operations are tested before the resources are inserted, therefore, the resources are not found. PETIT cannot identify this test case as being *failed (as expected)* as a result of both these operations preconditions being very permissive, as shown in listings 6.8 and 6.9. Since preconditions do not fail, PETIT classifies the tests as inconclusive.

```
1  "/players/{playerNIF}":
2    get:
3      summary: Return a player by NIF.
4      x-requires:
5      - T
6      x-ensures:
7      - T
```

Listing 6.8: YAML partial object for Player's API get player operation.

```
1  "/tournaments/{tournamentId}":
2    get:
3      summary: Return a tournament by ID.
4      x-requires:
5      - T
6      x-ensures:
7      - T
```

Listing 6.9: YAML partial object for Tournament's API get tournament operation.

The MOC order strategy not only can be used to test mutators in a failure scenario but also observers in the same scenario, as shown in the previous example.

Player's API mutator operations have the same test results as the previous execution – with MCO strategy. However, tournament's API test results do not show the operation responsible for checking whether a player is enrolled in a tournament classified as inconclusive, since, this time, neither the player nor the tournament exist. As such, both operation's preconditions fail and the test result is *failed (as expected)* and the operation's implementation classified as being according to the specification, i.e., *ok*.

## 6.3   Testing Observers

Testing tournaments' application with both OMC and OCM order strategies the test results are the same as the ones described in the previous section – section 6.2 – when testing it with MOC strategy. Both APIs have an inconclusive operation test and it happens to be the same ones – retrieving a player and a tournament –, for the exact same reasons.

Testing observers immediately before constructors, assuming constructors are implemented according to its specification, one should check if the previously inserted resources are, in fact, shown. Testing observers immediately after mutators, assuming mutators implementation is according to its specification, one should look for discrepancies on whether what was modified by the mutators is shown when testing observers. Hereupon, every single operation order strategy is equally useful to test observer operations.

## 6.4 Tournaments' Application: faulty scenario

As mentioned in the beginning of this chapter, there is the need to test PETIT in a faulty application in order to figure out if it is capable of finding out if a microservice's implementation is, in fact, according to its specification. This section's listings depict PETIT's output when executed only in verbose mode – *v* flag. Once more, the tournaments' application is used as a base example, and as such, several implementation errors are added to its implementation. The new implementation of tournaments' application features six different errors:

**Tournament Deletion** the specification states that if all preconditions hold then the microservice will return the tournament that was removed from the system. In this case, instead of returning the resource, the microservice returns null.

**Enrollment Deletion** the player is not disenrolled from the tournament.

**Tournament Insertion** the tournament is inserted with missing information.

**Tournament Update** the tournament supposed to be updated remains the same as it was before.

**Player Insertion** the player is not stored in the system. Listing 6.10 depicts PETIT's output in this scenario, executed with COM strategy. By checking the operation postcondition results, one can conclude that the player was not, in fact, stored in the system.

```
>> POST /players
 > Verifying Invariants                    :  OK
 > Generating Data                         :  OK
 > Verifying Preconditions                 :  OK
 > Performing Request                      :  OK
 > Response
 { "playerNIF": "259447224",
   "firstName": "PEbz NO_YPWtB8OuyOuDvWCu7AOMcI-PnWOzgRAmW",
   "lastName": "ffxY7 u__vJS1ObWfESY1JCEhkd5PPNEG",
   "address": "v58FjjkPCnB5etMka59kstZnuDYWx13rBNDVCRzJFmmJcKv",
   "email": "6_-_.9@g.B",
   "phone": "291956980",
   "tournaments": []
 }
```

```
 > Verifying Postconditions                  :  NOT OK
   > Failed:
     - response_code(GET /players/259447224) == 200
-----------------------------------------------------------------
POST /players                                 :  NOT OK
```

Listing 6.10: PETIT's test results for the faulty player insertion.

**Player Deletion** the wrong player gets deleted. Listing 6.11 shows PETIT result for this operation's test, when executed with CMO order strategy. This operation's specification states that it should retrieve the player that got deleted. However, by analysing PETIT's output one can see that the retrieved player was not the one supposed to be deleted, as shown by the second postcondition's results. The first postcondition states that after deletion, the player should not be found and, also fails because the wrong player got deleted.

```
>> DELETE /players/{playerNIF}
 > Verifying Invariants                       :  OK
 > Recycling Data                             :  OK
 > Verifying Preconditions                    :  OK
 > Performing Request                         :  OK
 > Response
 { "playerNIF": "100123123",
   "firstName": "ana",
   "lastName": "ribeiro",
   "address": "rua 1",
   "email": "ana@ana.ana",
   "phone": "999999999",
   "tournaments": [
      { "tournamentId": 1,
        "tournamentName": "Triwizzard Tournament 2020",
        "capacity": 3,
        "playerNumber": 0,
        "players": []
      }
   ]
 }
 > Verifying Postconditions                   :  NOT OK
   > Failed:
     - response_code(GET /players/158536692) == 404
     - response_body(this)==previous(response_body(GET /players/158536692)
-----------------------------------------------------------------
DELETE /players/{playerNIF}                   :  NOT OK
```

Listing 6.11: PETIT's test results for the faulty player deletion.

In order to find the relationship between operation order and error detection PETIT was subject to several tests. Table 6.1 depicts the tests' results. As seen in table 6.1, not every order strategy detects every error. By only analysing the table it may seem that PETIT is not very good when testing mutator operations. Considering only the failing cells, i.e. the ones with ✗, one can see that the error is not detected because the operation order is not suitable for testing mutators for their success scenario. In every single time PETIT did not detect an error on a mutator operation, the strategy chosen always tested mutators before constructors and, consequently, there was no sufficient data to find the implementation errors.

|                      | CMO | COM | MCO | MOC | OCM | OMC |
| -------------------- | --- | --- | --- | --- | --- | --- |
| Player Deletion      | ✓   | ✓   | ✗   | ✗   | ✓   | ✗   |
| Tournament Deletion  | ✓   | ✓   | ✗   | ✗   | ✓   | ✗   |
| Enrollment Deletion  | ✓   | ✓   | ✗   | ✗   | ✓   | ✓   |
| Player Insertion     | ✓   | ✓   | ✓   | ✓   | ✓   | ✓   |
| Tournament Insertion | ✓   | ✓   | ✓   | ✓   | ✓   | ✓   |
| Tournament Update    | ✓   | ✓   | ✗   | ✗   | ✓   | ✗   |

Table 6.1: Error detection in each order strategy.

## CONCLUSIONS AND FUTURE WORK

This chapter features this work's conclusions as well as the possible future improvements to PETIT and APOSTL.

### 7.1 Conclusions

PETIT – a**P**i t**E**s**TI**ng **T**ool – is developed with the purpose of automating the microservice testing process. Its implementation falls into black-box testing, more precisely, into the specification-based testing approach. As such, PETIT only needs the microservices' specification in order to be able to test them. Although these specifications have useful information, there is still the need to complement it with more information so the testing could be thorougher. APOSTL – **A**PI **Pr**O**perty **S**pecifica**T**ion **L**anguage – is developed for this purpose and, as the name implies, is a language developed to formally annotate APIs with properties that will, ultimately, constitute an API contract.

Nowadays the industry is dangerously migrating to microservice architectures without a reliable and automated process for effectively testing the software it is using. This thesis contributions work towards the mitigation this problem, contributing not only with a specification language purposely built to formally specify microservices' API contracts, but also with a testing tool capable of generating (non-redundant) test data, and automatically testing the microservices' implementation.

Several tests are conducted in order to ascertain whether PETIT's behaviour is according to what is expected. PETIT is tested against a correct and a faulty application. The test results on the correct application have shown that although PETIT's output concerning the whole specification is positive, there is still the need to analyse the entirety of the execution trace. This need arises from the fact that an operation should be tested for its every possible outcome. As shown in chapter 6, that is, usually, not the case with a single

PETIT execution. The tests conducted in the faulty application are positive, meaning PETIT is able to find every introduced error, when provided with the appropriate order strategy. The test results also shown that the order strategy parameter should be carefully considered when using PETIT.

To summarize, the contributions initially planned were successfully achieved. This work contributions are an API specification language developed to specify API contracts, an algorithm which automatically generates test data for microservices, based on their extended specification, and, finally, a tool integrating both of these features and automating the microservice testing process. However, the language, the algorithm, and the tool itself can be improved. At this stage, neither PETIT nor APOSTL are developed at their highest potential.

## 7.2 Future Work

As previously referred, both PETIT and APOSTL implementations have room for improvement. In the current implementation, PETIT is only able to test an operation once per execution. It is important that, in the future, PETIT is able to test operations several times during a single execution to, e.g., test numerical invariants such as the one depicted in listing 5.2. In PETIT's current implementation there is no way to test the previous invariant when the *capacity* property is greater than 1, since the operation responsible for inserting a tournament is not tested more than once, and every test data is deleted from the database when PETIT's execution is over, i.e., assuming deletion operations are implemented conforming to their specification.

PETIT should also be able to test each API operation independently. Currently, the only way a user can manipulate the operations being tested is by changing the API testing order – *r* flag – or the operation order strategy. Besides having control on the operation order, users should also have control on which operations are being, in fact, tested.

APOSTL's implementation can also be enhanced by improving expressiveness. This can be achieved by changing APOSTL's grammar in order to accept properties such as nested quantifiers, as described in section 5.1.4. APOSTL is a specification language that can be used with any API description language that supports being extended. Currently, PETIT only supports OAS but it can also support other common used description languages such as RAML [42] – RESTful API Modeling Language.